\documentclass{aa}  

\usepackage{graphicx}
\usepackage{txfonts}
\usepackage{hyperref}
\usepackage{wrapfig}
\usepackage{orcidlink}
\usepackage[labelfont=bf, justification=justified, format=plain]{caption}
\usepackage[justification=justified]{subcaption}
\usepackage{url}
\usepackage{wasysym}
\usepackage[normalem]{ulem}
\usepackage{xcolor}
\hypersetup{
    colorlinks=true,
    linkcolor=blue,
    filecolor=magenta, 
    pdftitle={Overleaf Example}, 
    pdfpagemode=FullScreen,    
    urlcolor=blue,
    citecolor=blue
}
\usepackage[T1]{fontenc}
\usepackage{siunitx}
\defcitealias{Valentini:2021}{V21}
\defcitealias{kocevski_rise_2024}{K25}
\defcitealias{DiMascia2021a}{DM21a}
\defcitealias{weingartner2001}{WD01}
\def\code#1{\texttt{#1}}
\usepackage{enumitem}

\definecolor{sgcolor}{HTML}{008000}
\definecolor{sccolor}{HTML}{ffbe33}
\definecolor{fdcolor}{HTML}{ff6600}

\def\msun{{\rm M}_{\odot}}
\def\zsun{{\rm Z}_{\odot}}
\def\lsun{{\rm L}_{\odot}}

\def\mum{\mu{\rm m}}

\def\angstrom{\textrm{A\kern -1.3ex\raisebox{0.6ex}{$^\circ$}}}

\begin{document}

   %\title{Is GN-z11 powered by a super-Eddington accreting massive black hole?: Extreme value statistics application}
   \title{Exploring the observed properties of JWST Little Red Dots with cosmological simulations and dust radiative transfer}

   \subtitle{}
   \authorrunning{Saksham Chandna et al.}
   \titlerunning{Exploring the observed properties of JWST LRDs with cosmological simulations and dust RT}
   \author{Saksham Chandna \fnmsep\thanks{\href{mailto:saksham.chandna@sns.it}{saksham.chandna@sns.it}} 
          \inst{1},
          Fabio Di Mascia\orcidlink{0000-0002-9263-7900}\inst{1},
          Simona Gallerani\orcidlink{0000-0002-7200-8293}\inst{1},
          Stefano Carniani\orcidlink{0000-0002-6719-380X}\inst{1},
          Pierluigi Rinaldi\orcidlink{0000-0002-5104-8245}\inst{2},
          Roberta Tripodi\orcidlink{0000-0002-9909-3491}\inst{3},
          Milena Valentini\orcidlink{0000-0002-0796-8132}\inst{4,5,6},
          Alessia Tortosa\orcidlink{0000-0003-3450-6483}\inst{7}, 
          Luca Zappacosta\orcidlink{0000-0002-4205-6884}\inst{3}
          %\fnmsep
          %\thanks{Just to show the usage
          %of the elements in the author field}
          }

   \institute{Scuola Normale Superiore, Piazza dei Cavalieri 7, I-56126 Pisa, Italy.
        \and Cosmic Frontier Center, University of Texas at Austin, Austin, TX 78712 Texas, USA.
        \and INAF - Osservatorio Astronomico di Roma, Via Frascati 33, I-00078 Rome, Italy.
        \and Dipartimento di Fisica dell'Università di Trieste, Sez. di Astronomia, Via Tiepolo 11, I-34131 Trieste, Italy.
        \and INAF - Osservatorio Astronomico di Trieste, via Tiepolo 11, I-34131, Trieste, Italy.
        \and INFN - Istituto Nazionale di Fisica Nucleare, Via Valerio 2, I-34127, Trieste, Italy.
        \and INAF OAS -Osservatorio di astrofisica e scienza dello spazio, Via Piero Gobetti, 93/3, 40129 Bologna BO, Italy.
        %Instituto de Estudios Astrof\'{\i}sicos, Facultad de Ingenier\'{\i}a y Ciencias, Universidad Diego Portales, Avenida Ejercito Libertador 441, Santiago, Chile
        %\and 
        %INAF – Osservatorio Astronomico di Trieste, via Tiepolo 11, \textcolor{red}{I-}34131 Trieste, Italy
        %\and
        %IFPU - Institute for Fundamental Physics of the Universe, Via Beirut 2, 34014 Trieste, Italy
        %\and
        %Astronomy Unit, Department of Physics, University of Trieste, via Tiepolo 11, \textcolor{red}{I-}34131 Trieste, Italy 
        %\and 
        %ICSC - Italian Research Center on High Performance Computing, Big Data and Quantum Computing
        %\and 
        %Dipartimento di Fisica, Sapienza, Università di Roma, Piazzale Aldo Moro 5, 00185 Roma, Italy
        %\and
        %Gemini Observatory, NSF’s NOIRLab, 670 N A’ohoku Place, Hilo, \textcolor{red}{Hawai'i 96720}, USA
             }

   \date{Received --; accepted --}

% \abstract{}{}{}{}{} 
% 5 {} token are mandatory 
\abstract
%
% context
%
%\begin{abstract}
{The James Webb Space Telescope (JWST) has uncovered a population of compact high-redshift ($z\gtrsim4$) sources, dubbed ``Little Red Dots'' (LRDs), characterised by a distinct V-shaped continuum (blue in the rest-frame UV, red in the optical). Their nature remains debated because broad-line AGN signatures coexist with non-AGN-like continua, marked by weak emission in the X-ray and mid--infrared.}
%Their physical nature remains debated because broad-line AGN signatures appear alongside non-AGN features, including weak X-ray and hot dust emission. 
%Their nature remains debated because clear AGN signatures often coexist with weak X-ray and hot-dust emission.}
%
% aims
%
%{We investigate which of the observed properties of LRDs can naturally emerge within cosmological zoom-in hydrodynamical simulations of massive AGN-host galaxies, and which require additional physical ingredients. In particular, we aim to identify the mechanisms responsible for the characteristic compact morphology and V-shaped spectral energy distributions of LRDs, as well as the missing physics required to explain their absent or weakened AGN signatures.}
{We investigate which properties of LRDs naturally emerge within simulations of massive AGN host galaxies and which require additional physical ingredients.}
%
% methods
%
{We post-process two cosmological zoom-in simulations of the same massive halo, one including black hole growth and AGN feedback (\emph{AGNfid}) and the other with only star formation and stellar feedback (\emph{SFonly}). We analyse snapshots spanning $z=6\text{--}9$ to probe different evolutionary stages. Using SKIRTv8, we develop a forward-modelling framework exploring various dust models, dust-to-metal ratios, and intrinsic AGN spectra. We generate spectra and mock JWST/NIRCam images to apply observational photometric and morphological LRD selection criteria.}
%
% results
%
{Not designed to model LRDs, our \emph{AGNfid} simulations naturally reproduce their properties. Specifically, we recover the V-shaped continuum, compact morphologies with rest-UV complexity, UV luminosities, attenuations, overmassive black holes, sub-Eddington accretion rates, and Balmer break strengths ($F_{4000}/F_{3600}\sim1\text{--}2$) typical of most LRDs. Spectral decomposition reveals that the rest-UV continuum is dominated by stellar emission, whereas the optical continuum stems from a dust-attenuated AGN. Dust geometry and the AGN-to-stellar spectral contribution drive the photometric differences between LRDs and their bluer counterparts (Little Blue Dots).}
%
% conclusions
%
{Our results suggest that LRDs do not necessarily represent a distinct class of objects. Instead, their defining properties can arise in massive AGN host galaxies from the interplay between stellar emission, an obscured central engine, and dust geometry. Super-Eddington accretion, dust-free gas columns, evolved stellar populations, and photoionisation modelling are required to determine whether this framework can reproduce the full diversity of the observed LRD population.}
%\end{abstract}
\maketitle
%
%-------------------------------------------------------------------

\section{Introduction}
Over the past few years, the \textit{James Webb Space Telescope} (\textit{JWST}) has revolutionised our understanding of galaxy formation and black-hole (BH) growth in the early Universe \citep{adamo2024billionyearsaccordingjwst, Matthee_2025}. Among its most remarkable discoveries is an abundant population of active galactic nuclei (AGN) at $z>4$, much more numerous than predicted by pre-\textit{JWST} models \citep{onoue_candidate_2023, Kokorev_2023, harikane2023, larson2023, Maiolino_2024, greene_uncover_2024}. Unlike the rare, luminous quasars ($M_{\rm UV}\lesssim-24$) traditionally identified at these redshifts, the majority of these newly discovered AGN are significantly fainter ($M_{\rm UV}\gtrsim-22$), probing a previously inaccessible regime of black-hole growth.

A substantial fraction ($\sim10-30\%$; \citealt{Hainline_2025}) of these AGN belong to a previously unknown class of compact, optically red sources, dubbed ``Little Red Dots'' (LRDs) \citep{labbe_population_2023, kokorev_2024, matthee_little_2024}. They peak in number density around $z\sim5$ before rapidly declining toward lower redshifts \citep{kokorev_2024, Tanaka_2024, kocevski_rise_2024, inayoshi_little_2025, Lin2026,  lin2026lrds2lowredshiftlittlered, Ma_2026}. However, recent studies caution that this apparent drop-off may be driven, at least in part, by restrictive photometric colour cuts and the limited depth of existing surveys, which preferentially select the brightest sources. Indeed, LRD-like systems have also been identified at low and intermediate redshifts \citep{Loiacono2026, rinaldi2026waytallytaleimpact, kapoor2026littlereddotsz2}.

LRDs are observationally characterised by compact, point-source-dominated morphologies in the rest-frame optical and distinctive V-shaped spectral energy distributions (SEDs), defined by a blue rest-frame UV continuum and a steep red optical continuum \citep{greene_uncover_2024, Perez2024, kokorev_2024, kocevski_rise_2024}. On average, this SED inflection is centered near the Balmer break \citep{setton_little_2024}, though individual sources exhibit spectral turn-overs extending to longer rest-frame wavelengths \citep{barro2026}.

Approximately $60-90\%$ \citep{barro2026} of the spectroscopically confirmed LRDs exhibit broad H$\alpha$ emission lines, together with prominent FeII complexes commonly associated with type-I AGN \citep{greene_uncover_2024, matthee_little_2024, Hviding_2025, Tripodi_2025a}. However, several other observational properties appear inconsistent with a standard unobscured AGN interpretation. Their relatively flat mid-infrared SEDs show little evidence for hot dust emission from an AGN torus \citep{Williams2023, Perez2024, Akins2024, Leung_2025, Wang_2025, Iani_2025, Setton_2025, Li_2025, Barro_2026}, while deep X-ray observations have yielded mostly non-detections \citep{yue_stacking_2024, ananna_x-ray_2024, maiolino_jwst_2025, tortosa2026xrayweaknesslittlered}, with rare individual exceptions \citep[e.g.,][]{hviding2026xraydotexoticdust, kocevski_rise_2024}. Furthermore, multi-epoch observations reveal little or no UV variability, unlike the stochastic variability commonly observed in AGN \citep{Furtak_2025, Kokubo_2025, Tee_2025, Zhang_2025}.

Another intriguing feature of the LRD population is the presence of strong Balmer breaks. In many systems, these are broadly consistent with evolved stellar populations similar to those inferred for post-starburst galaxies \citep{baggen_small_2024, Furtak2024, greene_uncover_2024, labbe_unambiguous_2024, kokorev_2024, Wang_2024, Wang_2025, ma_uncover_2024}. However, a small fraction ($\lesssim 10 \%$) of objects displays Balmer breaks significantly stronger than those that can be reproduced by conventional stellar population synthesis models \citep{perezgonzalez2026littlereddotsphotometric}. Notable examples include MOM-BH*1 \citep{naidu2025blackholestarreveals} and \textit{The Cliff} \citep{DeGraff2025}, whose extreme spectral discontinuities require additional physical mechanisms beyond normal stellar evolution. For these objects, proposed explanations include absorption by dense shells of neutral hydrogen in the excited $n=2$ state \citep{inayoshi_extremely_2025, ji_blackthunder_2025}, although the physical mechanism responsible for generating and sustaining such high column densities of excited neutral gas remains a matter of active debate.

The detection of spectroscopic AGN signatures alongside the absence of other canonical AGN indicators has made the physical nature of LRDs a central debate of the \textit{JWST} era. Taken together, their observed properties are difficult to reconcile within a single physical framework. Consequently, a broad range of theoretical models have been proposed, including dense nuclear star clusters \citep{baggen_small_2024}, black hole-stars \citep{DeGraff_BHstar_2025}, direct-collapse black holes (DCBHs) \citep{pacucci2026littlereddotsdirect}, and rapidly accreting massive black holes operating under super-Eddington scenarios \citep{chen2026supereddingtonaccretionblackholes, madau_maioliono2026}. Other frameworks suggest dense neutral-hydrogen envelopes capable of producing extreme Balmer absorption \citep{inayoshi2025criticalevaluationphysicalnature}, ionised cocoons surrounding supermassive black holes \citep{Rusakov_2026}, or hybrid stellar--AGN configurations \citep{kocevski_rise_2024}. However, these models typically explain features limited to specific subsets of the LRD population, which has proven to be highly heterogeneous \citep{perezgonzalez2026littlereddotsphotometric}.

The physical mechanism driving the characteristic V-shaped continuum varies depending on the assumed nature of the LRD. The steep rest-frame optical slope is often attributed to a dust-reddened AGN, a gas-enshrouded central engine, or an obscured, compact post-starburst stellar population \citep{labbe_population_2023, Barro_2023, volonteri_exploring_2025, setton_little_2024}. Dust-reddened scenarios face significant tension, as most LRDs remain undetected in the far-infrared, implying limited amounts of cold dust \citep{Akins2024, Casey_2025, Xiao_2025, Setton_2025}. However, the broad Balmer emission lines strongly favor an AGN-dominated scenario \citep{Kokorev_2023, Furtak2024, greene_uncover_2024}. Meanwhile, the origin of the blue rest-frame UV continuum remains highly debated. The proposed mechanisms include scattered or partially attenuated AGN emission \citep{ji_blackthunder_2025, Tripodi_2025, Torralba_warm_2026, pacucci2026littlereddotsdirect, deugenio2026blackthunderstrikestwicerestframe} or a relatively less attenuated UV stellar continuum escaping from the host galaxy \citep{DeGraff_BHstar_2025, Baggen_2026}.

From a morphological standpoint, the extreme compactness of these sources ($r_{\text{eff}} \lesssim 100\text{ pc}$ at $z\sim5$; \citealt{onoue_candidate_2023, labbe_population_2023}) strongly constrains pure stellar models. Reconstructing these structural profiles solely through stellar assemblies requires extreme stellar mass densities ($\gtrsim 10^4\text{--}10^5\,M_{\odot}\,\text{pc}^{-3}$) and velocity dispersions ($\sim 1000\text{ km/s}$) \citep{baggen_small_2024, ma_uncover_2024, Guia_2024, chisholm2026littlereddotsglobular} that are not observed locally and are fundamentally unsupported by dynamical models \citep{Hopkins_2010, Grudi_2019}. Nonetheless, high-resolution imaging increasingly reveals faint, extended rest-frame UV structures and close companion sources associated with a substantial fraction of the LRD population ($\sim 40\%$, increasing to $\gtrsim 80\%$ for ultra-luminous sources with $L_{5100} > 2 \times 10^{44} ~\rm erg ~s^{-1}$; \citealt{Baggen_2023,baggen2025resolvingcomplexmultiscalemorphology, Tanaka_2024, labbe_unambiguous_2024, Rinaldi_2025, rinaldi2026dotlrdlikenucleusheart, yanagisawa2026venusfaintlittlered}). These extended components may signify emergent host-galaxy stellar light, active mergers, or neighbouring galaxies that provide the critical Lyman-Werner flux ($11.2\text{--}13.6\,\text{eV}$) required by the DCBH interpretation \citep{Baggen_2026}.

For LRDs with spectroscopic confirmation of broad Balmer lines (typically H$\alpha$ or H$\beta$), single-epoch virial estimates yield black hole masses spanning $M_{\rm BH} \sim 10^{5\text{--}8}\,M_{\odot}$. The corresponding Eddington ratios based on broad-line luminosities typically range between $\lambda_{\rm Edd} \sim 0.1\text{--}1$ \citep{ananna_x-ray_2024, yue_stacking_2024}. To account for the overmassive nature of these central black holes, numerous frameworks invoke rapid, super-Eddington accretion regimes \citep{inayoshi2024birthrapidlyspinningovermassive, chen2026supereddingtonaccretionblackholes, chon2026rapidemergenceovermassiveblack}, which simultaneously helps explain their apparent X-ray weakness \citep{Pacucci_2024, madau_x-ray_2024, husko2025effectssupereddingtonaccretionfeedback, madau_maioliono2026}. Additionally, the stellar masses inferred from the broadband SED fitting and independent dynamical mass constraints consistently place LRDs significantly above the local $M_{\rm BH} - M_{\star}$ relations \citep{harikane2023, pacucci2023, maiolino_jwst_2025, Sun_2025, li2026dichotomynuclearhostgalaxy}.

Systematic modelling and estimation biases may also artificially drive this apparent overmassiveness \citep{Chen_2025, Li_overmassive_2025,roberts2026,ziparo2026}. In particular, frameworks incorporating electron or resonant scattering attribute line broadening to radiative effects rather than pure kinematics, reducing inferred virial masses \citep{naidu2025blackholestarreveals, Torralba_warm_2026, Rusakov_2026, Chang_2026}. However, \citealt{Juodbalis_2026} recently used resolved gas kinematics of the prototypical LRD Abell2744-QSO1 at $z=7.04$ to provide the first direct dynamical confirmation of its overmassive virial estimate. This result implies a critical evolutionary phase in which a central black hole dominates the host mass budget, consistent with rapid growth or heavy seed nature \citep{Inayoshi_2022, Hu_2022, Scoggins_2023}. %Radiative transfer modelling of these overmassive systems can directly verify if such extreme mass ratios inherently correlate with observed LRD features.

Any viable physical model for LRDs must also account for the surprising ubiquity of the population, given their high number densities of $\sim 10^{-4}\text{--}10^{-5}\,\text{cMpc}^{-3}$ across $z \sim 4\text{--}9$ \citep{matthee_little_2024, greene_uncover_2024, Akins2024, kocevski_rise_2024, zhuang2025nexusspectroscopiccensusbroadline}. Such high abundances severely challenge exotic formation scenarios, which must explain how these systems form so efficiently on a population scale. %Purely stellar interpretations invoking massive star-forming galaxies require masses of $\sim 10^{10\text{--}11}\,M_{\odot}$, severely violating the limits of standard structure formation beyond $z \gtrsim 7$ \citep{Wang_2024, Akins2024}. In the AGN framework, the UV luminosity function of these faint sources ($\Phi \sim 10^{-4}\text{--}10^{-5}\,\text{cMpc}^{-3}\,\text{mag}^{-1}$ for $-22 \le M_{\rm UV} \le -18$) is overabundant by an order of magnitude compared to ground-based quasar extrapolations \citep{Niida_2020, kokorev_2024, pizzati2025, rinaldi2026waytallytaleimpact}. Resolving this physical nature has far-reaching consequences for theories of early SMBH formation \citep{bogdan2023evidenceheavyseedorigin, Kovacs_2024}, host galaxy co-evolution \citep{Juod_balis_2024, maiolino_jwst_2025}, and cosmology \citep{Boylan_Kolchin_2023}. Therefore, combining cosmological simulations with radiative transfer provides a unique approach to test how naturally these LRD-like properties emerge within standard paradigms.

In this study, we compare the observed properties of LRDs with predictions from cosmological zoom-in simulations by \citealt{Valentini:2021} (hereafter \citetalias{Valentini:2021}), originally developed to study the formation of massive $z\sim6$ quasars \citep{fan2006, Matsuoka_2016, Matsuoka_2019, gallerani2017, banados2018, Vito_2019, pensabene2020}. We post-process the simulations using the dust radiative transfer code \code{SKIRT} \citep{Camps2015} %originally applied to the simulations of \citealt{Barai_2018})
and develop a pipeline to generate synthetic spectra and mock \emph{JWST}/NIRCam observations. This approach allows us not only to determine which observational properties of LRDs naturally emerge within a standard AGN-host galaxy framework, but also to identify the additional physical ingredients required to reproduce the diversity of the observed LRD population.

This paper is organised as follows. Sec.~\ref{allsims} details the hydrodynamical simulations and radiative transfer modelling. The subsequent mock observation pipeline is developed in Sec. \ref{mocks}. Sec. \ref{results} presents our results, including synthetic spectral energy distributions (Sec. \ref{res:SEDs}), compactness \citep{Labbe_2025} and photometric selection based on \citealt{kocevski_rise_2024} (hereafter \citetalias{kocevski_rise_2024}; Sec. \ref{res:LRDphoto}), JWST observational comparisons (Sec. \ref{res:observational}), and intrinsic physical properties (Sec. \ref{res:properties}). We interpret the nature of our simulated LRD population in Sec.~\ref{res:interpretation} and summarize our conclusions in Sec. ~\ref{summary}.
\vspace{-0.2cm}
\section{Numerical Model} \label{allsims}
We post-process the hydrodynamical cosmological zoom-in simulations developed by \citetalias{Valentini:2021} with Radiative Transfer (RT) calculations that include dust, following the procedure described in \citealt{DiMascia2021a} (hereafter \citetalias{DiMascia2021a}). We refer the reader to those works for detailed methodologies; here, we summarize key aspects and highlight differences in the hydrodynamic simulations and analysis framework relative to \citetalias{DiMascia2021a}. The main characteristics of the hydrodynamical simulations are described in Sec. \ref{hydrosim}, while the RT setup, including the assumed dust properties and the characteristics of the emission sources considered in the various post-processing runs, is presented in Sec. \ref{RTsim}.
\vspace{-0.2cm}
\subsection{Hydrodynamical simulations}\label{hydrosim}

We consider the \emph{SFonly} and \emph{AGNfid} runs presented in \citetalias{Valentini:2021}, which follow the evolution of a halo reaching a mass of $\sim 10^{12}$~M$_{\odot}$ at $z=6$. The former includes only star formation (SF) and stellar feedback, whereas the latter also accounts for BH seeding and quasar feedback. These simulations were performed with a non-public version of the TreePM (particle mesh) and SPH (smoothed particle hydrodynamics) code GADGET-3 \citep{valentini2017,valentini2019,valentini2020}, an upgraded version of the public GADGET-2 code \citep{springel2005}.

\vspace{-0.2cm}
\subsubsection{Initial conditions and resolution}
\label{subsec:Valentini_ICs}

The \code{MUSIC}\footnote{\code{MUSIC}–Multiscale Initial Conditions for Cosmological Simulations: \url{https://bitbucket.org/ohahn/music}.} software \citep{hahn2011} is used to generate the initial conditions for this simulation\footnote{A flat $\Lambda$CDM cosmology is assumed with the following parameters \citep{Planck:2016}: ${\Omega_{\rm M,0}= 0.3089}$, ${\Omega_{\mathrm \Lambda,0}= 0.6911}$, ${\Omega_{\rm B,0}= 0.0486}$, ${H_0 = 67.74~\rm{km~s}^{-1}~{\rm Mpc}^{-1}}$.}. At first, a dark matter (DM)-only simulation is run from $z=100$ to $z=6$, with a mass resolution of $m_{\rm DM} = 9.4\times 10^8~\msun$ for the DM particles inside a comoving volume of $(148~{\rm Mpc})^3$. Then, a halo as massive as $M_{\rm halo}=1.12 \times 10^{12}~\msun$ at $z=6$ is selected for the zoomed-in re-simulation with full hydrodynamical treatment of baryons. The zoomed-in simulation has the highest resolution particles with a mass of $m_{\rm DM}=1.55\times 10^6~\msun$ for DM, and $m_{\rm gas}=2.89 \times 10^5~\msun$ for gas. The gravitational softening lengths are $\epsilon_{\rm DM}=0.72$~ckpc\footnote{We use the following convention when indicating distances: a letter \emph{c} before the corresponding unit refers to \emph{comoving} distances (e.g. ckpc), while the letter \emph{p} refers to \emph{physical} units (e.g. pkpc). When not explicitly stated, we refer to physical distances.} and $\epsilon_{\rm bar}=0.41$~ckpc for DM and baryon particles, respectively.

\subsubsection{Sub-resolution physics}\label{sec:sub_phy}
The sub-grid physics implemented in \citetalias{Valentini:2021} can be summarised as follows:
\begin{enumerate}[itemsep=6pt]
    \item[$\bullet$]\textbf{Cooling, star formation and stellar feedback}:
    The simulations employ the MUlti Phase Particle Integrator (MUPPI) sub-grid model \citep{Murante2010, Murante2015, valentini2017, Valentini2018, valentini2019} to describe a multi-phase interstellar medium (ISM). The model includes metal-lines cooling, an $H_{2}$-based star formation criterion, thermal and kinetic stellar feedback, the presence of a UV background, and a model for stellar chemical evolution. Star formation is implemented using the stochastic model by \citep{Springel2003}, where a fraction of cold gas in molecular phase can turn into a star particle over its dynamical timescale, with a SF efficiency of $f_{\star} = 0.06$. Each star particle is described by a stellar population assuming the \citealt{Chabrier2003} initial mass function (IMF) and releases heavy elements (13 different metals, plus Hydrogen and Helium) through chemical enrichment channels, as described in \citealt{Tornatore2007}. Feedback energy is supplied in both thermal and kinetic forms by each star-forming multiphase particle to neighbouring gas particles, with coupling efficiencies of $f_{\rm fb, therm} = 0.2$ and $f_{\rm fb, kin} = 0.12$ , respectively \citep{Valentini2018,valentini2019}.
    %Should we add more description about the star formation and feedback?
    \item[$\bullet$]\textbf{Black holes seeding}: DM halos exceeding the threshold mass $M_{\mathrm {DM}}=1.48 \times 10^9~\mathrm{M}_\odot$ are seeded with a collisionless BH particle ($M_{\mathrm{BH, seed}} = 1.48 \times 10^5~\mathrm{M}_\odot$) at the gravitational potential minimum, provided no BH has already been seeded in the halo. This choice of seed mass is intended to mimic the direct collapse BH scenario, which predicts SMBH seeds with masses $M_{\rm BH, seed}\sim 10^4-10^6~\msun$ \citep{haehnelt1993, ferrara2014, mayer2019}. %As described in \citetalias{Chakraborty_2023}, 
    \item[$\bullet$]\textbf{BH accretion}: BHs grow through both gas accretion and BH-BH mergers. Gas accretion is described by the classical Bondi-Hoyle-Lyttleton accretion solution \citep{Hoyle1939, Bondi_Hoyle, bondi1952, Edgar2004}:
    \begin{equation}  \label{eq-Mdot-Bondi} 
        \dot{M}_{\rm Bondi} = \frac{4 \pi G^2 M_{\rm BH}^2 \langle\rho\rangle}{ \left(\langle c_{\rm s}\rangle^2 + \langle v\rangle^2\right) ^ {3/2}} , 
    \end{equation}
    where $G$ is the gravitational constant \citep{Springeletal2005}. The gas density $\langle\rho\rangle$, its sound speed $\langle c_{\rm s}\rangle$, and the BH velocity relative to the gas $\langle v\rangle$ are evaluated by kernel-weighted averaging over the SPH gas particles within the BH smoothing length. 
    Eq. \ref{eq-Mdot-Bondi} is used to estimate the contribution to the accretion rate from the cold and hot phase of the ISM separately \citep{Steinborn2015, valentini2020}. However, accretion from cold gas is additionally limited by the high angular momentum of the gas \citep[see][for details]{valentini2020}. The BH accretion rate is capped to the Eddington accretion rate.

    \item[$\bullet$]\textbf{BH repositioning and merging}: BH repositioning or \emph{pinning} is implemented: at each time-step BHs are instantaneously shifted towards the location of minimum gravitational potential within their softening length  \citep[as also done in e.g.][]{booth2009, schaye2015, Weinberger2017, Pillepich2018}. Furthermore, BHs can instantaneously merge if the following conditions are both satisfied: (i) their distance becomes smaller than twice their gravitational smoothing length; (ii) their relative velocity satisfies the following condition: $v_{\rm BH-BH} < 0.5 \ \langle c_{\rm s}\rangle$. The merged BH is placed at the position of the more massive progenitor BH.
    
    \item[$\bullet$]\textbf{AGN feedback}: A fraction of the accreted rest-mass energy is radiated away with a radiative efficiency $\epsilon_{\rm r}$, thereby providing a bolometric luminosity for a BH equal to:
    \begin{equation} \label{eq-Lr-BH} 
        L_{\rm bol} = \epsilon_{\rm r} \dot{M}_{\rm BH}c^2,
    \end{equation}
    where $c$ is the speed of light and $\epsilon_{\rm r} = 0.03$.  
    %Should we make comments about this value of radiative efficiency? and whether its valid for an LRD?
    A fraction $\epsilon_{\rm f} = 10^{-4}$ (tuned to match the normalisation of the BH to stellar mass relation at $z=6$ in \citetalias{Valentini:2021}) of the radiated luminosity $L_{\rm bol}$ is coupled thermally and isotropically to the surrounding gas. This AGN feedback energy is distributed to the hot and cold phases of multiphase gas particles within the BH smoothing volume \citep{valentini2020}.

\end{enumerate}
%So we do not expect any large variation in mass accretion rates of the LISA detectable systems post merger nd hence the effect of a merger cannot be determined from following the accretion rates of the BHs involved in a coalescence.. 
%Fig.\ref{Mdot} shows the variation of the mass accretion rates of the primary and secondary black holes before coalascence and the total accretion rate of the MBH resulting from the merger. We see that the mass accretion rate after the merger remains comparable to that of the primary (usually the heavier) BH and hence the effect of a merger cannot be determined from following the accretion rates of the BHs involved in a coalescence.

\subsection{Radiative Transfer calculations}\label{RTsim}
We post-process $23$ snapshots per simulation run across the redshift range $z=6\text{--}9$ for both the \emph{SFonly} and \emph{AGNfid} suites using %full continuum Radiative Transfer (RT).  To execute these calculations including dust, we use
\code{SKIRT}\footnote{Version 8, \url{http://www.skirt.ugent.be}.}, a Monte-Carlo RT code, designed to model continuum radiation fields in dusty media. The code accounts for dust absorption and scattering, as well as the subsequent re-emission of the absorbed energy in the infrared (IR) \citep[e.g.][]{Baes2003, Camps2015}. The code samples the source SED with a finite number of photon packets and follows their propagation through the specified simulation volume, accounting for the local dust properties. We exploit the flexibility of \code{SKIRT}, which enables us to adopt different dust grain size distributions, compositions, and dust-to-metal ratios, as well as different intrinsic SEDs for radiatively emitting stars and accreting BHs. In total, we perform a total of $22$ distinct post-processing runs, with distinct parameter combinations summarised in Table~\ref{sim_run_table}. 
%The code also includes several physical processes, such as dust stochastic heating and self-absorption. \FDM{I would remove this last sentence as you specify these later.}
For the RT calculations, we select a cubic region with a side length of $60$ physical kpc, centered on the center of mass of the most massive halo at each redshift, so as to include all galaxies associated with the system. 
%The dust models and its corresponding properties are described in detail in Section \ref{sec:dust_properties}, and the different SED characteristics adopted are explained in Section \ref{sec:SEDs}.
\begin{center}
    \begin{table}
        \centering
        \begin{tabular}{c c c c c c}
            \hline \\ 
            Hydro Run & AGN SED & $f_\text{d}$ & $a_{\text{min}}$ [$\mu m$] & Dust Model\\ [1ex] 
             \hline \hline
             \textit{SF\_only} & - & 0.1 & - & SMC\\
             \textit{SF\_only} & - & 0.1 & - & MW\\
             \textit{SF\_only} & - & 0.3 & - & SMC\\
             \textit{SF\_only} & - & 0.3 & - & MW\\
             \textit{SF\_only} & - & 0.1 & 0.1 & SMC\\
             \textit{SF\_only} & - & 0.3 & 0.1 & SMC\\
             \textit{AGN\_fid} & fiducial & 0.05 & - & SMC\\
             \textit{AGN\_fid} & fiducial & 0.05 & - & MW\\
             \textit{AGN\_fid} & fiducial & 0.1 & - & SMC\\
             \textit{AGN\_fid} & fiducial & 0.1 & - & MW\\
             \textit{AGN\_fid} & fiducial & 0.3 & - & SMC\\
             \textit{AGN\_fid} & fiducial & 0.3 & - & MW\\
             \textit{AGN\_fid} & fiducial & 0.1 & 0.1 & SMC\\
             \textit{AGN\_fid} & fiducial & 0.3 & 0.1 & SMC\\
             \textit{AGN\_fid} & UV-steep & 0.05 & - & SMC\\
             \textit{AGN\_fid} & UV-steep & 0.05 & - & MW\\
             \textit{AGN\_fid} & UV-steep & 0.1 & - & SMC\\
             \textit{AGN\_fid} & UV-steep & 0.1 & - & MW\\
             \textit{AGN\_fid} & UV-steep & 0.3 & - & SMC\\
             \textit{AGN\_fid} & UV-steep & 0.3 & - & MW\\
             \textit{AGN\_fid} & UV-steep & 0.1 & 0.1 & SMC\\
             \textit{AGN\_fid} & UV-steep & 0.3 & 0.1 & SMC\\
        \end{tabular}
        \caption{\code{SKIRT} post-processing runs. The columns indicate: (\textbf{first}) the post-processed hydrodynamical simulation run from \citetalias{Valentini:2021}, (\textbf{second}) the AGN intrinsic SED (no BHs in \textit{SF\_only} simulations), (\textbf{third}) the dust-to-metal ratio, (\textbf{fourth}) the minimum grain size in the modified grain size distribution (if it is not specified, no cut is applied, and the standard grain size distribution is considered), and (\textbf{fifth}) the dust model as in \citetalias{weingartner2001}.}
        \label{sim_run_table}
    \end{table}
\end{center}
\vspace{-1.2cm}
\subsubsection{Dust properties} \label{sec:dust_properties}
Dust is modeled in the computational domain by assuming a linear scaling with the gas metallicity\footnote{Throughout this paper the gas metallicity is expressed in solar units, using ${\zsun=0.013}$ as a reference value \citep{Asplund2009}.} \citep[e.g.][]{Draine2007}, according to a \emph{dust-to-metal ratio} $f_{\rm d}$ that quantifies the mass fraction of metals locked into dust:
\begin{equation} \label{eq:fd}
    f_{\rm d} = M_{\rm dust} / M_Z,
\end{equation}
where $M_{\rm dust}$ is the dust mass and $M_Z$ is the total mass of all the metals in each SPH gas particle in the hydrodynamical simulation (see Sec. \ref{sec:sub_phy}). Gas particles hotter than $10^6$~K are assumed to be dust-free because of thermal sputtering \citep[e.g.][]{Draine1979}. 

We modify the total dust mass in our RT calculations by modifying $f_{\rm d}$. This parameter is poorly constrained for high-redshift sources both from observations and theoretical models \citep{Nozawa2015, Wiseman2017}, and therefore, is treated here as a free parameter. For our $z\geq 6$ sample, we adopt a baseline dust-to-metal ratio of $f_{\rm d} = 0.3$ (calibrated to the Milky Way) alongside lower alternatives. This range is motivated by theoretical models where dust formation efficiency scales with metallicity as galaxies evolve \citep{Asano2013, Aoyama2017}. %, though this behavior at early epochs remains heavily debated \citep{Ferrara2016}.
Crucially, a low overall dust content is strongly favored by recent observations of LRDs, which consistently feature far-infrared non-detections \citep[e.g.,][]{Labbe_2022, Akins2024, Xiao_2025, Setton_2025, Casey_2025} and exhibit minimal dust attenuation in their narrow emission lines \citep[e.g.,][]{Deugenio2025, Nikopoulos2025, Ji2026, Lin2026, ivey2026}. We therefore explore relatively low values of the dust-to-metal normalisation parameter $f_{\rm d}$. Specifically, we adopt three different values: (i) a MW-like value, $f_{\rm d} = 0.3$, (ii) a smaller value, $f_{\rm d} = 0.1$, for general high-redshift galaxies, and (iii) an even smaller value, $f_{\rm d} = 0.05$, to consider extremely dust-poor systems. 
% Nikopoulos paper important to argue that AGN is dust obscured but the galaxy not so much.
%Check out Wang et al 2024 and Ma et al 2025 cause theyu argure they need steep extinction curves for the steep rise.

For the dust distribution within the \texttt{SKIRT} RT framework, the simulation data is mapped onto a 3D cubic computational domain with $L = 60$~kpc. The dust geometry is structured using an octree grid—an adaptive mesh refinement (AMR) technique that hierarchically subdivides cells to accurately resolve high-density regions. We utilise up to ten levels of refinement for the dust distribution in the octree grid, corresponding to a maximum spatial resolution of $\approx 60$~pc, comparable to the baryonic resolution of the hydrodynamical simulations ($\epsilon_{\rm bar}=59$ ppc at $z = 6.0$, see Sec. \ref{subsec:Valentini_ICs}). 
%\SC{how do we justify using the same refinement for all redshifts?}

The grain size distribution and chemical composition of dust for early epoch galaxies is still an open question, since the origin and nature of dust at high redshift are highly debated \citep{Valiante2009, Stratta2011, Asano2013b, Hirashita2015, Hirashita2019}. Some studies involving high-redshift quasars and GRBs favour a Small Magellanic Cloud (SMC) type extinction curve \citep{Zafar2011, Zafar2018, Hjorth_2013}. Recently, however, researchers have used flatter/grayer attenuation curves to obtain the LRD "V-shaped" continuum spectra \citep{Li_2025, Chen_dust_2025, madau_maioliono2026}. MW-like extinction curves are flatter than SMC-like curves, producing stronger overall attenuation but weaker UV reddening for a given dust column. Therefore, we consider both SMC and MW-type dust compositions and grain size distributions, using the results of \citealt{weingartner2001} (hereafter \citetalias{weingartner2001}). 

Additionally, for both dust models implemented in \textsc{SKIRT}, the minimum grain size is set to $a_{\rm min}=0.001~\mu$m for silicate and carbonaceous grains, and to $a_{\rm min}=3.548\times10^{-4}~\mu$m for polycyclic aromatic hydrocarbons (PAHs). However, several observational studies have suggested that AGN-host galaxies may be deficient in small dust grains, based on the observed flat attenuation curves in these systems \citep{Gaskell_2004, Czerny2004, Gallerani2010}. This interpretation is also supported by numerical simulations \citep{DiMascia2021b} and theoretical models of dust evolution \citep{Nozawa2015, Hirashita2019}. Motivated by these findings, we additionally consider a modified version of the \citetalias{weingartner2001} dust model, imposing a minimum grain size of\footnote{The modified \citetalias{weingartner2001} dust model featuring a large-grain cutoff of $a_{\rm min}=0.1\,\mu\text{m}$ is utilised exclusively for our SMC-type extinction runs.  As noted by \citealt{DiMascia2021b}, the MW and SMC grain-size distributions are very similar for grains larger than $0.1~\mum$.} $a_{\rm min}=0.1~\mu$m. For all dust models, the grain size distribution of each dust component (silicates, graphitic grains, and PAHs) is represented using five size bins. In the SMC model, no PAH component is included.

Finally, dust self-absorption is included in the RT calculations, where dust grains reabsorb the IR photons emitted by the dust itself, which is critical in highly optically thick environments. The local dust grain temperatures and corresponding emissivities are computed self-consistently by enforcing strict energy balance between the total local radiation field and the thermal re-emission. The calculations are iterated until the integrated dust IR luminosity changes by less than $3\%$ between successive iterations. %Non-local thermal equilibrium (NLTE) corrections to dust emission are neglected, as is heating by the cosmic microwave background (CMB), since both effects are expected to be negligible. 
%\SC{Why?? How do I show this since this was not checked? IT seems many are comparable with the CMB temperatures} \FDM{You are right, this sentence is not well motivated and these assumptions were more appropriate for a different study. Regarding the CMB temperature, at z=9, it is around 28 K; I am not sure what is the temperature distribution within the objects considered in the analysis, but I am expecting a small fraction of dust cells with a temperature lower than that, so that should be ok (we can check for some cases though). Regarding the NLTE, it is more complicated, because it affects the PAH emission in all the simulations with MW-type dust. However, this part of the spectra shouldn't be relevant for any of our conclusions I think, so we should be safe. Overall, given that this is quite technical, we might omit this entire sentence and answer the referee if they asks about it.}

\subsubsection{Stellar and AGN radiation} \label{sec:SEDs}

Stellar and BH particles are treated as point sources of radiation, with their positions taken from the hydrodynamical simulation snapshots at each redshift. Each stellar particle represents a simple stellar population (SSP), i.e. a population of stars with the same age and metallicity. The emitted radiation is assigned according to the stellar population synthesis models of \citealt{Bruzual2003}, using the mass, age, and metallicity of each stellar particle to determine its spectral energy distribution (SED).

For BHs, we adopt the AGN spectral energy distribution introduced in \citetalias{DiMascia2021a}, which is described by a composite power-law:
\begin{equation}\label{AGN_SED_eq2}
     L_\lambda = c_i \ \left(\frac{\lambda}{\mu{\rm m}}\right)^{\alpha_i} \ \left(\frac{L_{\rm bol}}{\lsun}\right) \ \lsun \ {\mum}^{-1},
\end{equation}
where $i$ identifies the spectral interval considered. The coefficients $c_i$ are obtained by enforcing continuity between adjacent spectral segments and depend on the adopted slopes $\alpha_i$ (see Table~2 of \citetalias{DiMascia2021a} for their numerical values). For our RT calculations, we consider two intrinsic SED models for the AGN, with distinct UV spectral slope ($\alpha_{\rm UV}$): \emph{fiducial} ($\alpha_{\rm UV} = -1.5$) and \emph{UV-steep} ($\alpha_{\rm UV} = -2.3$), as described in \citetalias{DiMascia2021a}. 

The SED is normalised to the BH bolometric luminosity, computed using Eq. \ref{eq-Lr-BH}. The accretion disc and dusty torus ($\lesssim 10$ pc) are not resolved in our simulations. The AGN SED described by Eq. \ref{AGN_SED_eq2} includes emission from the accretion disc but does not include a hot dusty torus component. We will further discuss this point in Sec. \ref{summary}. 
%\SG{please add the discussion about no torus in low luminosity AGN. Resolved}

% Read more about hot-dust deficient AGN in Iani et al, and some LRDs have hot dust in Li. 

The radiation field is sampled using a grid composed of $200$ logarithmically spaced bins, covering the \emph{rest-frame} wavelength range $[0.1-10^3]$ $\mum$. Shorter wavelengths are not considered, as the RT calculations do not include hydrogen absorption for photons with $\lambda < 912 ~\AA$. For each wavelength bin, $10^6$ photon packets are emitted distributed over each radiation source. Radiation from photon packets exiting the computational volume is collected along six orthogonal viewing directions corresponding to the faces of the RT computational cube.
%Therefore, we are able to obtain six RT datacubes for each hydrodynamical simulation snapshot, with a field-of-view of $60 \times 60$ kpc ($1024 \times 1024$ pixels), and a wavelength axis of $200$ logarithmic bins.

\section{Mock Observations} \label{mocks}
%\SG{General comment for Saksham. Please make an effort to keep each figure as close as possible to the text discussing it.}
Each post-processed snapshot produces six synthetic spectral data cubes along lines of sight perpendicular to the faces of the simulation volume, assuming an observer located at a luminosity distance corresponding to the redshift of the snapshot. Each synthetic observation has a field of view (FoV) of $60~\text{kpc} \times 60~\text{kpc}$, sampled on a $1024 \times 1024$ pixel grid, yielding a spatial resolution of $58.59~\text{pc}$, consistent with that of the hydrodynamical simulation (i.e. $\epsilon_{\rm bar} \sim 58.6$~pc at $z=6$). In total, we obtain 3036 independent lines of sight for our analysis.

\begin{figure*}[h]
    \centering
    \includegraphics[width = \linewidth]{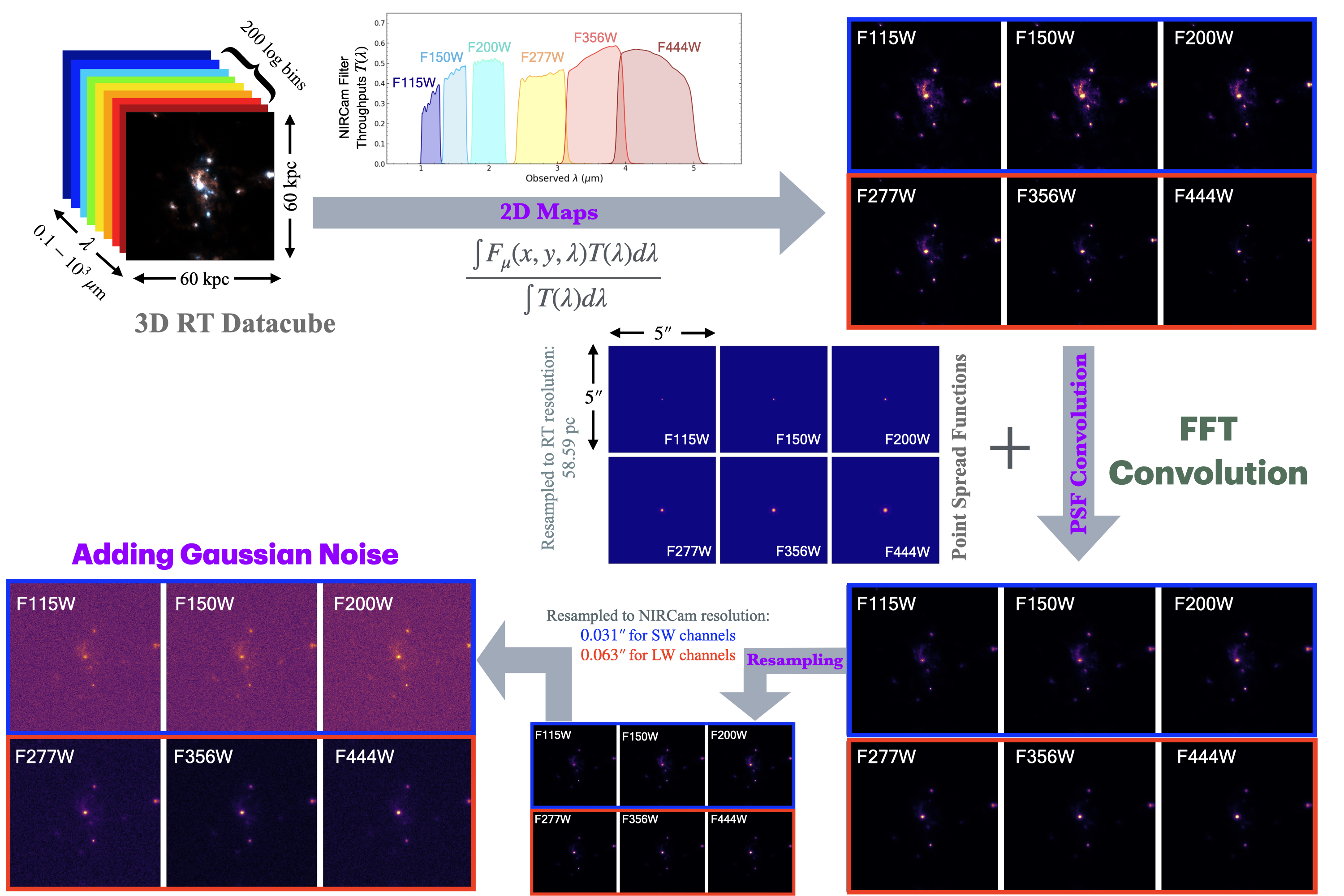}
    \caption{Schematic overview of the pipeline used to produce mock photometric maps from the radiative transfer outputs. For visualisation, the colour scales are saturated above fixed fractions of the peak flux, such that all values exceeding the threshold are assigned the same colour. The intrinsic 2D maps are capped at $1\%$ of the maximum flux, while the convolved and noise-added maps are capped at $50\%$ of the maximum flux, improving the visibility of extended structures in the presence of bright central pixels.}
    \label{fig:flowchart}
\end{figure*}

We further process synthetic data cubes to generate mock \emph{JWST} observables. Fig.~\ref{fig:flowchart} summarizes the mock image-generation pipeline. Each step is described in detail below.

\begin{enumerate} [itemsep=6pt]
    \item \textbf{2D Maps}: The 3D radiative transfer (RT) data cubes are convolved along their wavelength axis with the \emph{JWST}/NIRCam photometric filter transmission curves\footnote{The \emph{JWST}-NIRCam throughput curves are publicly available on the \textbf{\emph{JWST} User Documentation: NIRCam Filters}, \url{https://jwst-docs.stsci.edu/jwst-near-infrared-camera/nircam-instrumentation/nircam-filters}}. The resulting cubes are then integrated over wavelength to produce 2D filter maps for each object and each line of sight. For this work, we generate 2D filter maps for the NIRCam F115W, F150W, F200W, F277W, F356W, and F444W filters.

    \item \textbf{PSF Convolution}: 
    We import the point-spread functions (PSFs) for the NIRCam filters listed above from the \code{STPSF}\footnote{Python package \textbf{\code{STPSF} Documentation}, \url{https://stpsf.readthedocs.io/en/latest/index.html}} python package that provides us with simulated PSFs for \emph{JWST} instruments. These PSFs are oversampled by a factor of $4$ and include realistic detector effects (e.g. geometric distortion, charge transfer effects, inter-pixel capacitance) ensuring good agreement with observed \emph{JWST} PSFs. Our 2D filter maps have a finer spatial sampling than the native NIRCam pixel scale ($0.031"$  for short wavelength (SW) channels from $0.6 - 2.3$ $\mum$, and  $0.063"$ for long wavelength (LW) channels from $2.4-5.0$ $\mum$). We therefore use the oversampled PSFs, which are first resampled to match the pixel scale of our 2D maps ($58.59~\rm pc$ per pixel) and then convolve the corresponding filter maps with the resampled PSFs using a Fast Fourier Transform (FFT).

    \item \textbf{Resampling}: After PSF convolution, we resample the 2D maps to the native pixel scale of the corresponding \emph{JWST}/NIRCam channel, thereby generating the final mock \emph{JWST} photometric maps. Since small-aperture fluxes and source compactness depend directly on pixel size, this step enables a direct comparison with real observations.

    \item \textbf{Adding Gaussian Noise}: To model the observational noise, we adopt the $10\sigma$ sensitivity limits corresponding to a $10\,\mathrm{ks}$ integration time for each \textit{JWST}/NIRCam photometric filter\footnote{The assumed sensitivity is consistent with the wider pointing of the JADES Medium survey \citep{eisenstein2023}, and are based on the Exposure Time Calculator (ETC) v5.0, which is publicly available on the \textbf{\emph{JWST} User Documentation: NIRCam Sensitivity}, \url{https://jwst-docs.stsci.edu/jwst-near-infrared-camera/nircam-performance/nircam-sensitivity}.}. Assuming the noise to be Gaussian, the signal-to-noise ratio can be written as $\text{S/N} = F^{\text{conv}}_{\nu}/\sigma\sqrt{N_{\rm ap}}$, where $F^{\text{conv}}_{\nu}$ is the convolved flux measured within a $2.5$ pixel aperture for a point source, $\sigma$ is the standard deviation of the Gaussian noise distribution, and $N_{\rm ap}$ is the number of pixels contained within the $2.5$ pixel aperture.

    First, we correct the $10 \sigma$ sensitivity limits provided by \emph{JWST} for PSF convolution using the encircled energy (EE) corrections available for each NIRCam filter\footnote{The Encircled Energy (EE) corrections for each NIRCam photometric filter are available on \textbf{\emph{JWST} User Documentation:}, \url{https://jwst-docs.stsci.edu/jwst-near-infrared-camera/nircam-performance/nircam-point-spread-functions\#NIRCamPointSpreadFunctions-Encircledenergy}}. Using the EE values corresponding to a $2.5$ pixel aperture ($0.08''$ for SW channels and $0.16''$ for LW channels), we determine the value of $F^{\text{conv}}_{\nu}$ in the above expression for $\text{S/N}$. We then adopt $\text{S/N}=10$ together with the corresponding value of $N_{\rm ap}$ to estimate the value of $\sigma$ for each photometric filter. These $\sigma$ values provide a simple estimate of the Gaussian noise to be added to our mock photometric maps and allow us to assess whether our objects are detectable. The resulting photometric maps including noise are illustrated in Fig.~\ref{fig:flowchart}.
\end{enumerate}

%\subsection{Photometric Maps} \label{sec:photometric}\FDM{I think this subsection can be aggregated to the previous one. There are only 2 paragraphs and without much content} 

The resulting F444W photometric maps are utilised to identify and select LRD candidates. To avoid systems in which the BH has only recently been seeded, we restrict our analysis to AGN-host galaxies located at the center of the most massive halo in the \emph{AGNfid} simulation. For each selected system, we then identify the brightest object in the F444W map within a projected radius of $5$ kpc from the center of the most massive halo and use it for all subsequent analyses. For consistency, we apply the same criterion to the star-forming galaxies in the \emph{SFonly} simulation.
%The parameters for the \citetalias{Valentini:2021} simulations are also tuned to ensure that the central BH and its host galaxy are in agreement with the $M_{\rm BH} - M_{\star}$ relation at $z=6$ \citep{pensabene2020}. 
 %Since the simulations involve a lot of merging systems, there is a possibility of multiple contaminating objects. We will focus our analysis only on the compact objects classified in Section \ref{res:compact}, as the brightest object is likely to be an isolated system if it satisfies the compactness criterion. 

We extract photometric fluxes using a $0.4''$ diameter aperture, consistent with the photometric aperture sizes commonly adopted in the literature ($0.3"-0.5"$ used in \citealt{kokorev_2024, Hviding_2025}). Hereafter, all fluxes extracted are corrected for the flux losses introduced by the PSF by multiplying them by the corresponding Encircled Energy (EE) correction for the relevant NIRCam filter and photometric aperture. 

\section{Results} \label{results}
 
In this section, we present the results obtained from the simulated SEDs and mock \emph{JWST} observations. %In Section~\ref{res:SEDs}, we analyse the simulated SEDs extracted from the post-processed data cubes after masking the regions immediately surrounding our target sources. In Section~\ref{res:LRDphoto}, we apply to the mock photometric maps the compactness criterion proposed by \citealt{Labbe_2025} together with the photometric LRD selection criteria by \citetalias{kocevski_rise_2024}.

%The following subsections place the selected sources in the context of current observations. In Section~\ref{res:observational}, we analyse the multi-wavelength observational properties of our simulated LRD candidates and compare them with those commonly associated with the observed LRD population. In Section~\ref{res:properties}, we examine the the underlying physical properties of the selected LRDs to determine whether they differ systematically from the broader simulated galaxy population. Finally, in Section~\ref{res:interpretation}, we investigate the physical and geometric mechanisms that give rise to LRD-like observational properties in our simulations.
\subsection{Synthetic Spectral Energy Distributions} \label{res:SEDs}
\begin{figure}[h]
    \centering
    \includegraphics[width = \linewidth]{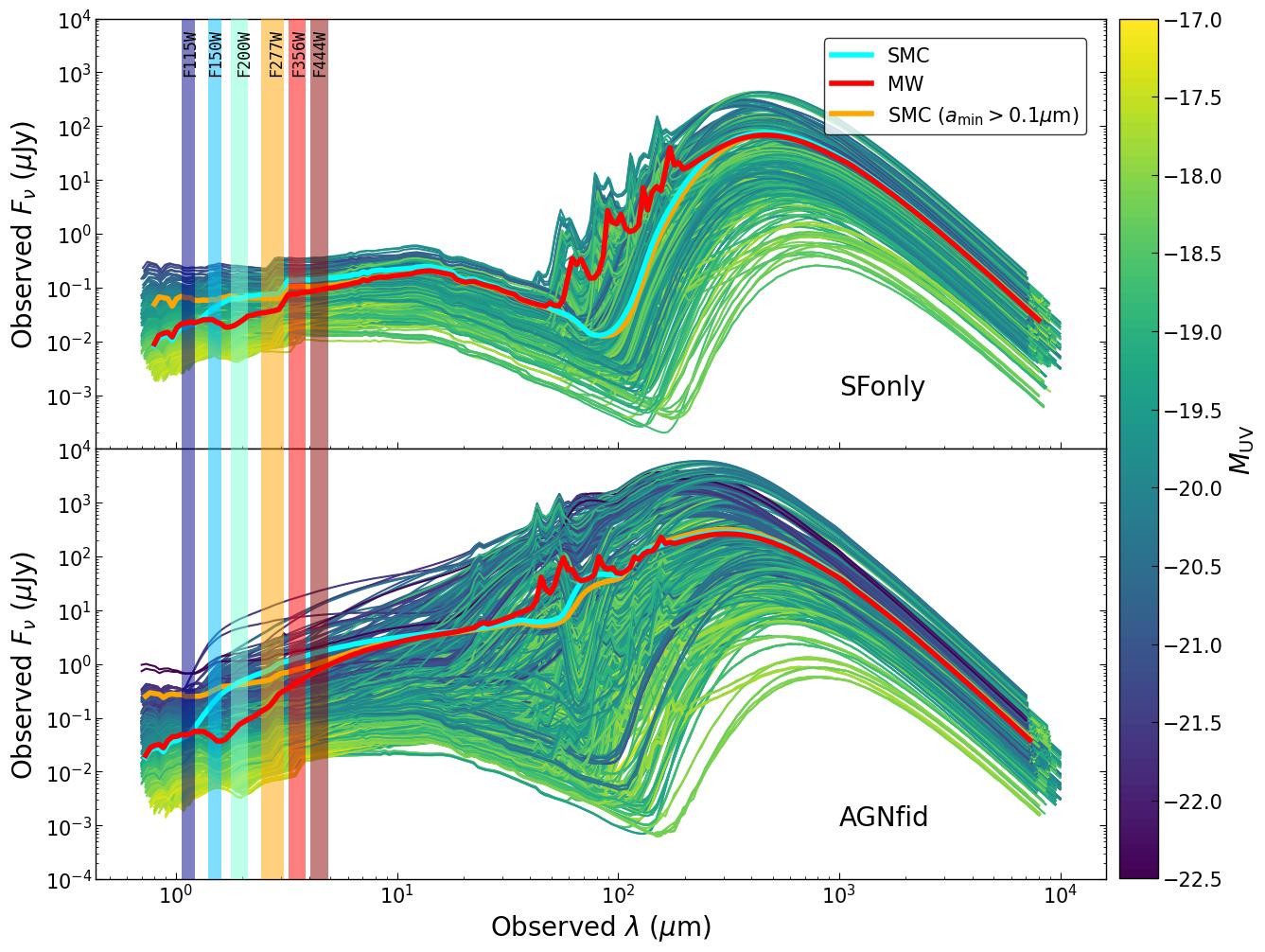}
    \caption{Synthetic spectral energy distributions (SEDs) extracted within an aperture diameter of $D_{\text{ap}} = 0.4''$ centered on the brightest F444W source within the central $5$~kpc of the most massive halo. Panels display results for the \emph{SFonly} (\textbf{top}) and \emph{AGNfid} (\textbf{bottom}) simulation suites. Shaded vertical bands indicate the bandpasses of the JWST/NIRCam filters utilised in our photometric selection pipeline (see Sec. \ref{res:LRDphoto}). Individual spectra are colour-coded by their emergent absolute UV magnitude ($M_{\text{UV}}$) to demonstrate the population variance in UV luminosity. To highlight the differences in dust absorption and thermal re-emission across different dust extinction models from \citetalias{weingartner2001}, a representative case is highlighted in distinct colours: with SMC-like extinction (cyan), MW-like extinction (red), and the modified SMC featuring a small-grain cutoff of $a_{\text{min}} > 0.1\,\mu\text{m}$ (orange).}
    \label{fig_obs:1}
\end{figure}

Fig.~\ref{fig_obs:1} shows the SEDs extracted using a $D_{\rm ap} = 0.4''$ diameter aperture applied directly to the 3D spectral data cubes generated by our RT calculations (see Sec.~\ref{RTsim}). Each simulated object is observed along six perpendicular lines of sight, with each line of sight treated as an independent mock observation of an LRD candidate. The two panels show the results for the \emph{SFonly} and \emph{AGNfid} simulation suites separately.

The \emph{SFonly} and \emph{AGNfid} suites exhibit comparable flux levels in the rest-frame UV (ranging from $10^{-3} \rm \mu Jy$ to $0.5-1 \rm \mu Jy$), whereas the \emph{AGNfid} spectra become systematically brighter from the rest-frame optical through the mid-infrared (MIR; rest-frame $\lambda \sim 1.4$--$15~\mu$m) and far-infrared (FIR; rest-frame $\lambda \sim 15$--$10^3~\mu$m). This additional emission originates from the central AGN. More specifically, the brightest \emph{SFonly} systems reach MIR flux densities of up to $\sim 0.5~\mu$Jy, compared to $\sim 10 \mu$Jy for the most luminous \emph{AGNfid} spectra. In the FIR, the thermal dust emission peaks at $\sim 0.4~$mJy and $\sim 6~$mJy for the \emph{SFonly} and \emph{AGNfid} suites, respectively.
These differences can be explained as follows. Because the AGN resides at the centre of the galaxy, where the gas and dust densities are the highest, its primary UV and optical radiation experiences substantially stronger line-of-sight attenuation than the more extended stellar component. Consequently, the emergent rest-frame UV continuum remains largely dominated by stellar light, producing broadly overlapping $M_{\rm UV}$ distributions between both simulation suites with the exception of a few low-obscuration sightlines where unattenuated AGN emission drives $M_{\rm UV}$ to brighter values. %Consequently, the emergent rest-frame UV continuum remains dominated by stellar emission, yielding similar $M_{\rm UV}$ values in the two simulation suites. 
However, at longer wavelengths, the dust-attenuated AGN continuum gradually emerges and overtakes the stellar emission, producing the enhanced MIR luminosity. The absorbed stellar and AGN radiation is subsequently re-emitted by the surrounding dust in the FIR. Because of the additional heating supplied by the AGN, the peak of the thermal dust emission is shifted toward shorter wavelengths in the \emph{AGNfid} simulations than in the \emph{SFonly} suite. For the brightest systems, the FIR emission peaks at rest-frame wavelengths of $\sim 400~\mu$m in the \emph{SFonly} suite, compared to $\sim 200~\mu$m in the \emph{AGNfid} simulations. A detailed analysis of this behavior based on spectral decomposition is presented in Sec.~\ref{res:interpretation}.

To illustrate the impact of the adopted dust model, Fig.~\ref{fig_obs:1} highlights three representative radiative transfer calculations performed on the same simulated system: a standard SMC extinction curve (cyan), a MW extinction curve (red), and a modified SMC model with a minimum grain size of $a_{\rm min}=0.1\,\mu$m (orange). While all three models exhibit the same overall transition from the attenuated stellar and AGN continuum to thermal dust re-emission in the FIR, they differ in several distinctive spectral features. Although broad $9.7\,\mu\mathrm{m}$ silicate absorption ($\lambda_{\mathrm{obs}}\simeq70\text{--}100\,\mu\mathrm{m}$ at $z\sim6$) is present across all models, the MW model uniquely displays prominent PAH emission bands at rest-frame $3.3$, $6.2$, $7.7$, $8.6$, $11.3$, and $12.7\,\mu\mathrm{m}$. These features are produced by stochastically heated polycyclic aromatic hydrocarbon (PAH) molecules and are absent in the SMC dust mixtures, producing a much smoother MIR continuum. In the rest-frame UV-to-optical regime, the SMC curve produces a steeper continuum slope and lower overall optical extinction than the MW model. %, driven by a deficit of large grains that reduces attenuation at longer wavelengths.} 
Finally, imposing a minimum grain size of $a_{\rm min}=0.1\,\mu$m suppresses the population of small grains responsible for the strongest UV absorption. As a result, the modified SMC model exhibits reduced attenuation and less reddening at extreme-UV wavelengths than the standard SMC extinction curve.
 %Finally, the modified grain size cut leads to reduced absorption in the extreme UV range due to a lack of small grains, resulting in less UV reddening than in the prototypical case of SMC-like extinction.
%In our modelling, these features disappear when shifting from the MW mixture to our other two setups: the standard SMC model and the modified SMC model with grain size cutoff ($a_{\rm min} = 0.1\,\mu\text{m}$) following \citetalias{weingartner2001}. In both SMC-based configurations, the lack of a transiently heated PAH population causes the mid-infrared emission to smooth out into a continuous continuum baseline. \FDM{The last two sentences do not add anything}
%\SG{do we need to write this? I cannot see a V-shape from visual inspection...} \FDM{I agree; also having all the SEDs together does not help} Visual inspection reveals that many lines of sight naturally produce the classical "V-shape" profile: a blue or flat continuum across the short-wavelength bands (F115W, F150W, F200W) turning over into a steep, red power-law slope at longer wavelengths (F277W, F356W, F444W). In our \emph{AGNfid} models, this V-shape arises because the rest-frame UV is dominated by escaping star-forming light from the host galaxy, while the rest-frame optical is dominated by the emerging power-law of the dust-reddened AGN core (see Section \ref{ref}).

\subsection{LRD Photometric Selection} \label{res:LRDphoto}
To identify LRD candidates in our simulated sample, we apply the observational selection criteria from recent surveys to our mock \emph{JWST} photometric maps. Our goal is to assess whether our simulated galaxies naturally reproduce the observational properties of LRDs. Specifically, we require our mock sources to satisfy both the photometric ``V-shape'' selection introduced by \citetalias{kocevski_rise_2024} and the compactness criterion proposed by \citealt{Labbe_2025}. The two criteria are discussed separately below.

\paragraph{\textbf{Spectral V-Shape}} We first apply the photometric LRD selection proposed by \citetalias{kocevski_rise_2024} to the mock photometry extracted within a $D_{\rm ap}=0.4''$ aperture. This selection is based on the characteristic ``V-shape'' continuum observed in LRDs and is quantified through the rest-frame UV and optical continuum slopes.
The continuum spectral index $\beta$ (defined through $f_\lambda \propto \lambda^\beta$) is calculated across wavelength windows positioned blueward and redward of the rest-frame Balmer break. Following \citetalias{kocevski_rise_2024}, we determine these slopes by performing a $\chi^2$ minimisation fit to the standard linear magnitude--wavelength relation:
\begin{equation}
m_i=-2.5(\beta+2)\log(\lambda_i)+c,
\end{equation}
where $m_i$ represents the AB magnitude in the $i$-th filter with pivot wavelength $\lambda_i$. 
The filter combinations used to determine the rest-frame UV ($\beta_{\rm UV}$) and optical ($\beta_{\rm opt}$) slopes are adjusted according to redshift. For systems at $z<8$, we use the F115W, F150W, and F200W filters to estimate $\beta_{\rm UV}$, and the F277W, F356W, and F444W filters to estimate $\beta_{\rm opt}$. For systems at $z\ge8$, the wavelength coverage shifts to F150W, F200W, and F277W for $\beta_{\rm UV}$, and F356W and F444W for $\beta_{\rm opt}$.

Our RT post-processing explicitly models continuum processes and does not include nebular line emission associated with the gas phase. Consequently, synthetic photometry is unaffected by line contamination (such as [O III] + H$\beta$ or H$\alpha$ + [N II] entering filters like F356W and F444W). We therefore apply the photometric selection directly to the underlying continuum shape\footnote{The $\beta_{\rm UV}>-2.8$ criterion was introduced by \citetalias{kocevski_rise_2024} to exclude Galactic brown dwarf contaminants. Although we retain it for consistency with the observational selection, it does not affect our simulated sample.}:
\begin{enumerate}
    \item[(i)] $-2.8 < \beta_{\text{UV}} < -0.37$
    \item[(ii)] $\beta_{\text{opt}} > 0$.
\end{enumerate}

The $\beta_{\rm opt} > 0$ criterion adopted by \citetalias{kocevski_rise_2024} is designed to select sources with very red rest-frame optical continua, although the precise location of this threshold is somewhat arbitrary. Recent studies have shown that extending the selection to $\beta_{\rm opt} < 0$ reveals a population of similarly compact but optically bluer sources \citep{Hainline_2025, rinaldi2026waytallytaleimpact, billand2026littlereddotsreally, barrufet2026orientationevidencedistinctphysical}. These objects, often referred to as ``Little Blue Dots'' (LBDs; e.g., \citealt{scholtz2026littleredbluedots}), may represent less obscured counterparts of the canonical LRD population. We therefore include both populations in the following analysis while focusing primarily on LRDs; a detailed investigation of the LBD population will be presented in a separate work.

%\textbf{While the \citetalias{kocevski_rise_2024} selection criteria successfully identify the canonical red-optics population, strict cuts on $\beta_{\rm opt} > 0$ restrict the sample to the extreme red end of compact sources. Recent empirical studies emphasize that relaxing this optical slope constraint unveils a broader, intrinsically related population of compact sources that share similar morphological properties but display bluer rest-frame optical continua ($\beta_{\rm opt} < 0$) \citep{Hainline_2025, rinaldi2026waytallytaleimpact, billand2026littlereddotsreally, barrufet2026orientationevidencedistinctphysical}. Often termed ``Little Blue Dots'' (LBDs; e.g., \citealt{scholtz2026littleredbluedots}), these objects may represent lower-attenuation, less obscured, or orientation-variant counterparts to classic LRDs. To avoid omitting this intrinsically compact, bluer subpopulation, we explicitly track sources satisfying $\kappa < 1.7$ and $-2.8 < \beta_{\rm UV} < -0.37$ but failing the $\beta_{\rm opt} > 0$ threshold, alongside primary LRD candidates throughout our analysis.}

\paragraph{\textbf{Compactness}} We then evaluate whether our mock sources satisfy the compactness criterion\footnote{We caution that flux extraction from fixed apertures has been found to introduce up to 30-40\% of noise \citep[e.g.,][]{rinaldi2026waytallytaleimpact}. The resulting candidate fractions are therefore likely biased low and should be regarded as conservative lower limits. A morphology-based analysis would be required to quantify and correct for this incompleteness.} proposed by \citealt{Labbe_2025}:
\begin{equation}
    \kappa = \frac{f_{\text{F444W}}(D = 0.4'')}{f_{\text{F444W}}(D = 0.2'')} < 1.7,
\end{equation}
where $f_{\text{F444W}}(D = 0.4'')$ and $f_{\text{F444W}}(D = 0.2'')$ represent the fluxes extracted from circular apertures of diameters $D = 0.4''$ and $D = 0.2''$, respectively, centered on the brightest pixel. 
%Using the mock NIRCam F444W imaging maps generated in Section \ref{mocks}, we evaluate this compactness metric for the most luminous sources residing within a 5~kpc radius of the central massive halo's core. This spatial constraint effectively builds a clean sample of central compact objects while avoiding contamination from unassociated line-of-sight structures within the photometric mask.

%The threshold of $\kappa < 1.7$ is mathematically calibrated to isolate sources that remain unresolved within the PSF of the NIRCam F444W filter. %For these point-like systems, virtually all of the integrated stellar and AGN flux is captured within the primary $D_{\rm ap} = 0.4''$ photometric aperture. For methodological self-consistency, this identical $0.4''$ aperture is utilised for spectral extraction from our 3D RT datacubes.

%Fig. \ref{fig:selection} shows the distribution of compact sources is explored in the rest-frame spectral index plane. Notably, none of the pure star-forming galaxies in the \emph{SFonly} run satisfy the point-source morphology criterion, whereas $154$ out of $2,208$ independent lines of sight within the AGNfid suite are successfully classified as compact. 

\paragraph{\textbf{Detection}} Following the procedure outlined in Sec. \ref{mocks}, the background noise levels ($\sigma$) derived are used to compute the signal-to-noise ratio ($\text{S/N}$) for each NIRCam band. We classify any system with $\text{S/N} < 5$ in any of the six primary NIRCam selection filters as a non-detection. Furthermore, we apply an additional detection threshold of $\text{S/N}_{\text{F444W}} > 12$ following \citetalias{kocevski_rise_2024} to isolate robustly detected sources.

\begin{figure*}[h] 
    \centering
    \includegraphics[width = \linewidth]{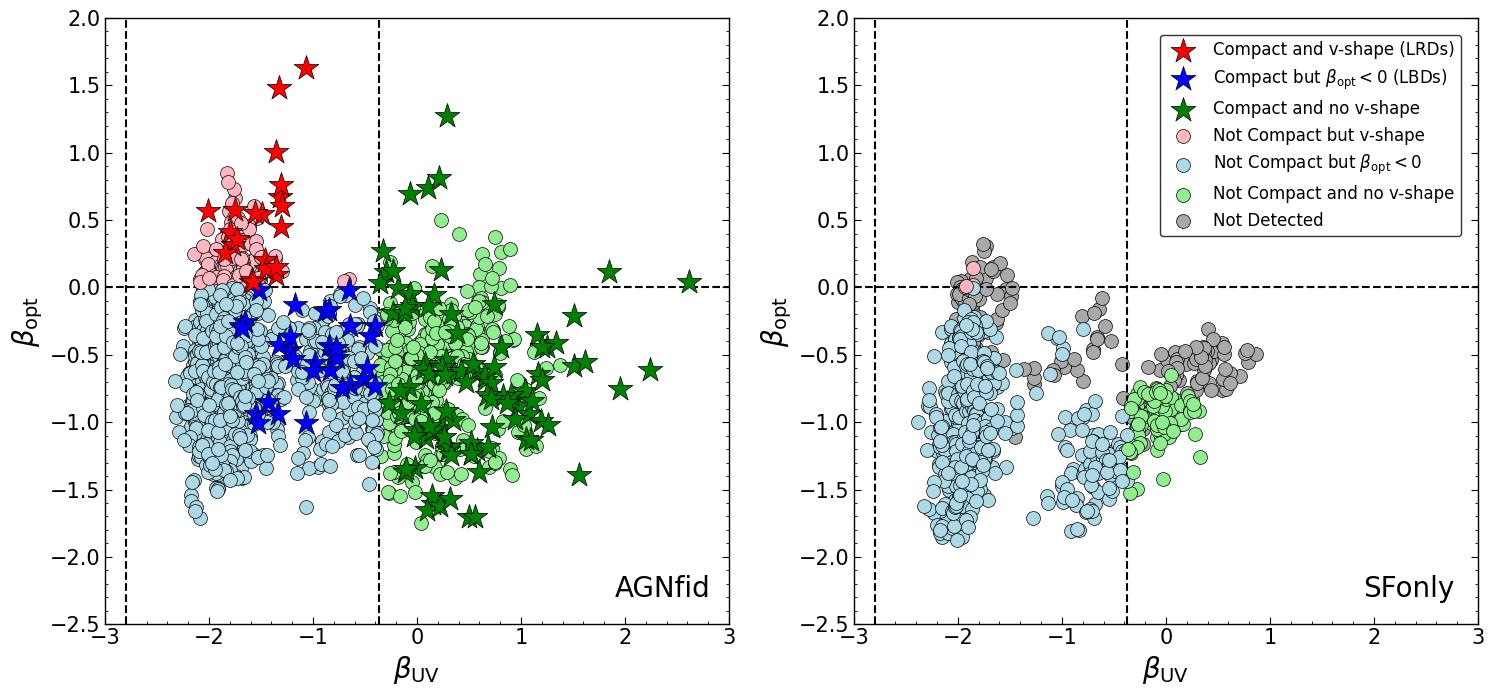}
    \caption{Distribution of synthetic sources in the rest-frame spectral index plane ($\beta_{\rm UV}$ vs. $\beta_{\rm opt}$) extracted from the \emph{AGNfid} suite (\textbf{left panel}) and the \emph{SFonly} suite (\textbf{right panel}). The horizontal dashed line at $\beta_{\rm opt}=0$ separates sources with red optical continua ($\beta_{\rm opt}>0$) from those with blue optical continua. The vertical dashed lines mark the boundaries of the rest-frame UV selection: $\beta_{\rm UV}=-0.37$ denotes the upper limit for blue UV colours, while $\beta_{\rm UV}=-2.8$ represents the lower cutoff adopted in empirical surveys to exclude Galactic brown dwarf contaminants. In both panels, star markers denote compact sources satisfying $\kappa<1.7$ (see Sec.~\ref{res:LRDphoto}), while circles indicate extended sources. Red symbols represent primary LRD candidates, satisfying both the compactness and full photometric (v-shape) criteria. Blue symbols denote compact sources with blue UV colours ($\beta_{\rm UV} \le -0.37$) that fail the optical cutoff ($\beta_{\rm opt} < 0$), while green symbols denote compact sources failing the UV selection. Pink symbols represent extended sources meeting the photometric criteria. Light-blue and light-green symbols identify extended sources meeting only the UV criterion or neither criterion, respectively. Gray symbols fall below the adopted detection threshold (Sec.~\ref{mocks}).
    %Red symbols indicate sources satisfying both the compactness and photometric selection criteria, and therefore constitute our simulated LRD candidates. Blue symbols denote sources satisfying the compactness and the $\beta_{\rm UV}$ criterion but not the $\beta_{\rm opt}$ lower limit. Green symbols denote sources satisfying only the compactness criterion, while pink symbols identify sources satisfying only the photometric criterion. Light-blue symbols represent sources satisfying only limits on $\beta_{\rm UV}$, while light green satisfy none of the criterion. Gray symbols correspond to sources falling below the adopted detection threshold (see Sec.~\ref{mocks})
    }
    \label{fig:selection}
\end{figure*}

Fig.~\ref{fig:selection} summarizes the outcome of applying the photometric selection and compactness criteria to our mock observations, showing the distribution of sources in the $\beta_{\rm UV}$--$\beta_{\rm opt}$ plane for both the \emph{AGNfid} and \emph{SFonly} simulation suites. This figure shows that both AGN-host and purely star-forming galaxies can reproduce the characteristic ``V-shaped'' continuum. %The physical origin of this feature, however, differs between the two populations, as further discussed in Sec. \ref{res:interpretation}. \FDM{I would add at least one sentence to say what is the physical origin and then refer to the discussion section.} 
Applying photometric selection alone, $227$ out of $2208$ independent lines of sight in the \emph{AGNfid} suite ($\sim 10\%$) satisfy the adopted colour criteria. A smaller fraction ($\sim 3\%$) of the \emph{SFonly} sample ($28$ out of $828$ sight lines) also satisfies these criteria. However, only $\sim 7\%$ of the AGN-host galaxies satisfy the compactness criterion, whereas none of the star-forming galaxies does\footnote{As discussed in the Introduction, purely stellar models can reproduce the compactness of LRDs only at very high stellar mass densities ($\gtrsim 10^{4}\,\mathrm{M}_{\odot}\,\mathrm{pc}^{-3}$), far exceeding those found in our simulations ($\sim 50\,\mathrm{M}_{\odot}\,\mathrm{pc}^{-3}$).}. Consequently, combining the photometric and structural selection criteria, the final sample of simulated LRD candidates represents only $\sim1\%$ of the total \emph{AGNfid} sources. 

Table~\ref{perc_table} summarizes how different post-processing configurations (dust models and intrinsic AGN spectra) affect both the fraction of compact sources and the final LRD selection. The adopted dust model has the strongest impact. SMC-type dust produces the largest fraction of compact systems because its weaker rest-frame optical attenuation allows the unresolved AGN continuum to dominate the F444W emission more readily than in the MW model. However, none of these systems satisfies the full LRD selection, as the steeper UV extinction curve of the SMC model reddens the rest-frame UV continuum beyond the adopted colour cuts. Conversely, all LRD candidates are associated with MW-type dust, whose flatter UV attenuation preserves a blue UV continuum while still producing the required red optical slope. The modified SMC grain-size distribution ($a_{\rm min}>0.1\,\mu$m) performs poorly in both compactness and LRD selection. Although removing the smallest grains reduces the UV attenuation, the larger grains increase the overall optical extinction and efficiently scatter the AGN radiation, suppressing compact nuclear emission.%\SC{Check the last point with decomposed spectra} 

The dust-to-metal ratio also strongly affects the selection. Increasing $f_{\rm d}$ progressively reduces the fraction of compact systems because the central AGN becomes increasingly attenuated relative to the more extended stellar component. The dependence of the LRD fraction is less straightforward, however, reflecting the complex interplay between dust geometry, wavelength-dependent attenuation, and the relative contributions of the stellar and AGN continua.

The intrinsic AGN SED has a more modest, yet non-negligible, impact. The fiducial template ($\alpha_{\rm UV}=-1.5$) produces both a larger fraction of compact sources and a higher LRD selection efficiency than the UV-steep model ($\alpha_{\rm UV}=-2.3$), owing to its higher intrinsic rest-frame optical flux and consequently steeper UV-to-optical continuum.

\begin{figure}[h] 
    \centering
    \includegraphics[width = \linewidth]{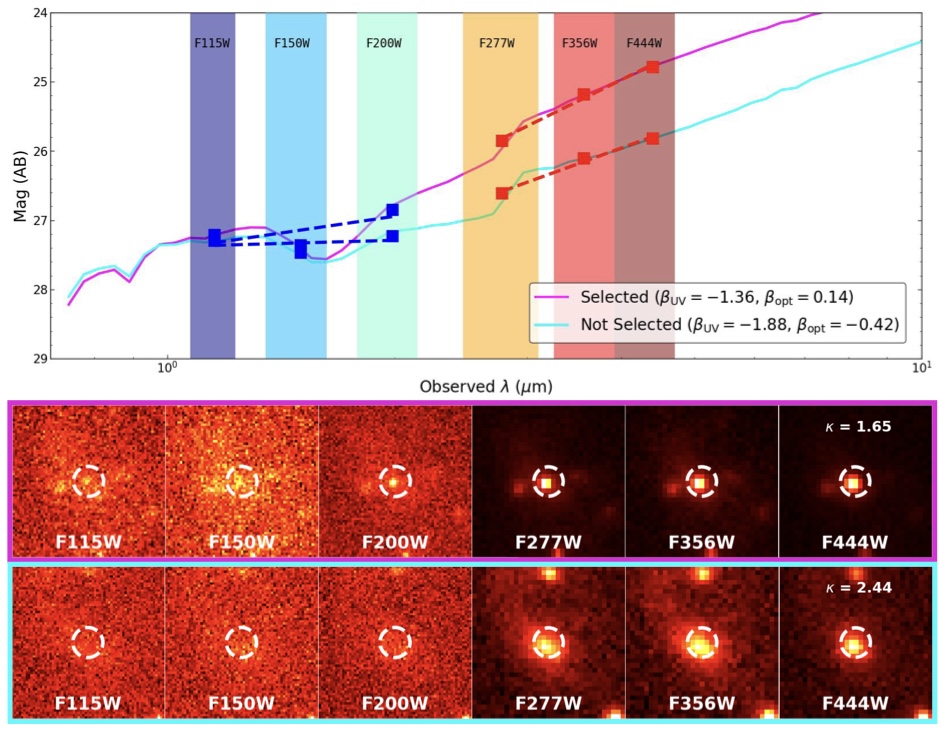}
    \caption{LRD selection and morphological compactness for two lines of sight of a simulated object at $z = 6.4$. The \textbf{top} panel shows synthetic spectral energy distributions (in AB magnitudes): one that satisfies the selection criteria (magenta) and one that lacks the characteristic V-shape (cyan). Shaded vertical bands mark the \emph{JWST}/NIRCam filters from the \citetalias{kocevski_rise_2024} selection, with squares indicating the extracted mock photometry (Sec.~\ref{mocks}). Blue and red dashed lines show power-law fits to the UV ($\beta_{\mathrm{UV}}$) and optical ($\beta_{\mathrm{opt}}$) slopes, respectively. The \textbf{bottom} panels display the corresponding NIRCam stamps for the selected (magenta outline) and unselected (cyan outline) sightlines, with white dashed circles indicating the extraction apertures and $\kappa$ stating the F444W compactness ratio. This comparison shows how changing the line of sight for the same object can alter its spectral slopes and compactness, driving it into or out of selection.}
    \label{fig:los_comp}
\end{figure}

Finally, line-of-sight variations also strongly affect both the inferred spectral slopes and morphological compactness of a system at a given snapshot. As illustrated in Fig. \ref{fig:los_comp}, increased obscuration along a specific line of sight can attenuate the central red AGN component in the longer-wavelength channels. This flattens the optical slope ($\beta_{\mathrm{opt}}$) while allowing extended emission to dominate the light profile, driving the object out of LRD selection under both colour and compactness criteria (see Sec. \ref{res:interpretation} for a detailed physical decomposition).

To summarise, our analysis provides the probability that a single massive halo satisfies observational LRD selection across its evolutionary history and viewing directions. By sampling a single cosmological zoom-in simulation across multiple snapshots and lines of sight, our mock dataset directly captures how orientation and evolutionary state regulate LRD observability. Consequently, the quoted $\sim 1\%$ selection fraction reflects the selection probability of this specific massive system under different viewing conditions, rather than the intrinsic occurrence rate or expected cosmic abundance of LRDs within the general galaxy population.
\begin{center}
    \begin{table}[]
        \centering
        \begin{tabular}{c c c c c c}
            \hline \\
            Dust Model & Compact (\%) & LRD (\%)\\ [1ex] 
            \hline \hline
              SMC & 73.4 & 0\\
              MW & 24 & 100\\
              Modified $a_{\rm min} > 0.1 \mum$ SMC & 2.6 & 0\\ [1ex]
            \hline \hline \\
            $f_{\rm d}$ & Compact (\%) & LRD (\%) \\ [1ex]
            \hline \hline
             0.05 & 75.3 & 72.7\\
             0.1 & 24 & 27.3\\
             0.3 & 0.7 & 0\\ 
            \hline \hline \\
            AGN SED & Compact (\%) & LRD (\%) \\ [1ex]
            \hline \hline 
             fiducial & 75.3 & 86.4\\
             UV-steep & 24.7 & 13.6\\
        \end{tabular}
        \caption{Fraction of simulated \emph{AGNfid} systems meeting the compactness and/or final LRD selection criteria for each post-processing parameter value (\textbf{left column}): adopted dust model (\textbf{top panel}), dust-to-metal ratio ($f_{\rm d}$; \textbf{middle panel}), and intrinsic AGN SED (\textbf{bottom panel}). The "Compact" column reports the percentage of the morphologically compact sub-sample possessing the corresponding model feature in the left column. Similarly, the "LRD" column reports the percentage of the fully selected LRD sub-sample possessing that feature.}
        \label{perc_table}
    \end{table}
\end{center}

\subsection{Comparison with JWST data} \label{res:observational}
In this section, we compare the \textit{JWST} observational properties of our simulated LRD candidates with those of the observed LRD population. Since our radiative transfer calculations do not include emission from a dusty torus (see Sec.~\ref{discussion}), the predicted MIR and FIR properties should be interpreted with caution. We therefore restrict our main comparison to the UV--NIR regime and present a preliminary comparison in MIR and FIR in Appendix~\ref{res:MIRI} and Appendix~\ref{res:fir_limits}. %We focus on directly observable quantities that can be measured from imaging and photometric data, enabling a direct comparison with available multi-wavelength observations.
%In this section, we investigate the multi-wavelength observational signatures commonly associated with the empirical LRD population. Our analysis focuses strictly on directly measurable quantities that can be determined from observations without relying on underlying physical model assumptions. First, in Section \ref{res:morph}, we examine the structural morphologies and small-scale environments of our sources using mock imaging maps. In Section \ref{res:MIRI}, we characterize the behavior of our simulated spectra within the \emph{JWST}/MIRI bandpasses. We then contrast our synthetic SEDs against observed far-infrared and submillimeter upper limits in Section \ref{res:fir_limits}. Finally, we evaluate how our simulated sample compares directly with the observed LRD population by analyzing their rest-frame UV brightness distributions in Section \ref{sec:uv} and quantifying their Balmer break strengths in Section \ref{res:balmerbreak}.

\subsubsection{Morphology} \label{res:morph}
\begin{figure*}[h]
    \centering
    \includegraphics[width = 0.98\linewidth]{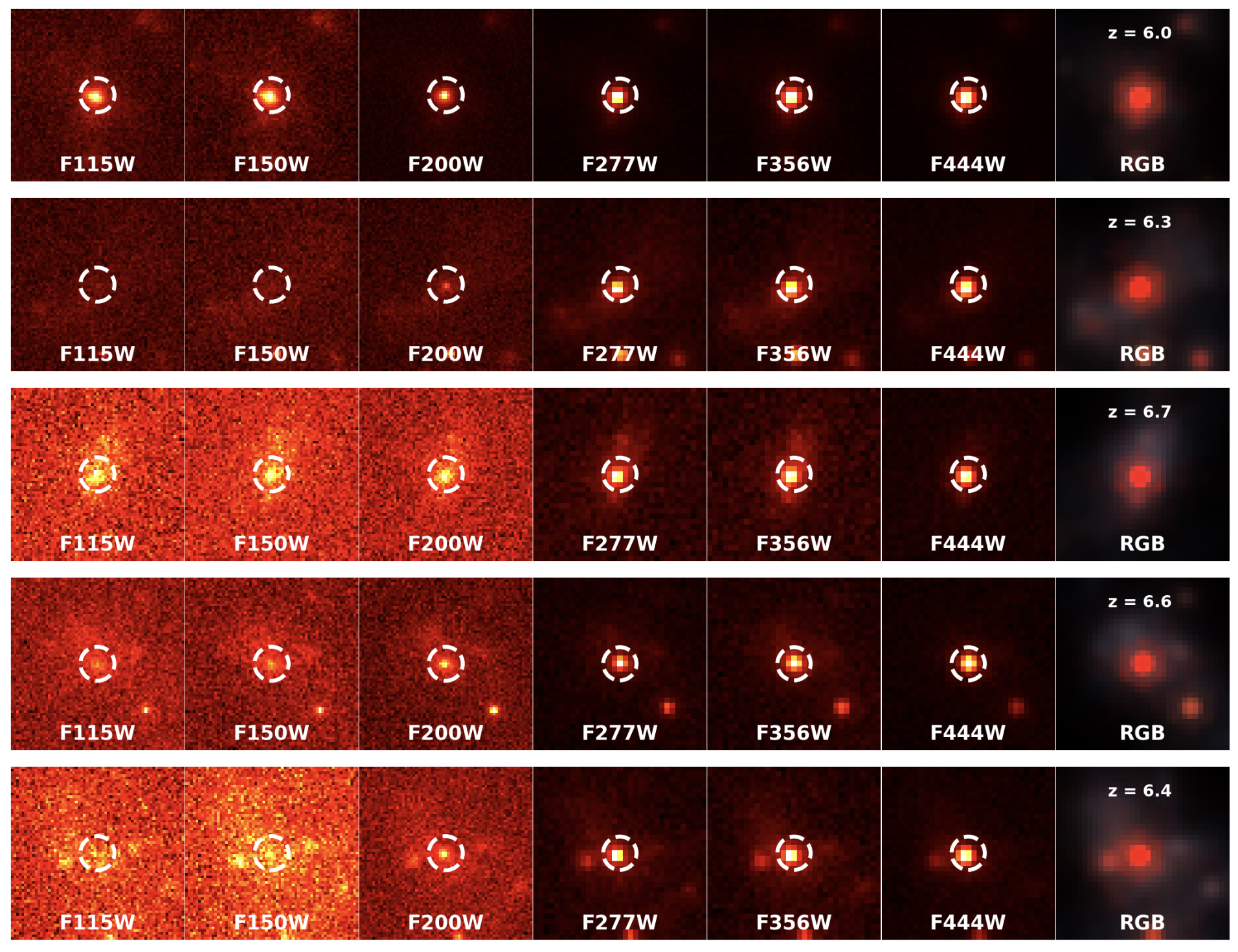}
    \caption{Mock \emph{JWST}/NIRCam image stamps and corresponding three-colour RGB composites generated using the post-processing procedure described in Sec. \ref{mocks} for a representative subset of the simulated LRDs selected in Sec. \ref{res:LRDphoto}. The redshift of each source is indicated in the RGB panel. Each panel shows a $2'' \times 2''$ field of view ($\sim 11\times 11$ physical kpc at $z=6.4$), centered on the most luminous source within the central massive halo in the NIRCam F444W band. The dashed white circle overplotted on each single-filter panel represents the $0.4''$ diameter photometric aperture ($D_{\rm ap} = 0.4''$, corresponding to $\sim 2$ physical kpc at $z=6.4$) used for flux extraction. Gaussian background noise is visually prominent in the SW filters (\textbf{left columns}), but less apparent in the LW filters (\textbf{middle-right columns}) due to the larger dynamic range imposed by the linear colour scale. The rightmost column shows RGB composites combining the F444W (R), F200W (G), and F115W (B) filters.}
    \label{fig_obs:stamps}
\end{figure*}
%\SG{check the reference to the different rows}

In Fig.~\ref{fig_obs:stamps}, we present mock \emph{JWST}/NIRCam images along with three-colour RGB composites\footnote{The RGB composites are generated following \citealt{Lupton_2004} by mapping the F444W, F200W, and F115W filters to the red, green, and blue channels, respectively.% We apply an arcsinh stretch with $Q=8$ to compress the dynamic range of bright nuclear cores and a linear scale parameter $\text{stretch}=0.1$ on the normalised flux arrays to enhance low-surface-brightness host features.
} of five representative simulated LRD candidates (see App. \ref{app:allmaps} for the complete atlas in Fig. \ref{fig:allmaps}). %\SG{Add this, when you show the atlas in the appendix.} 
This figure illustrates a broad diversity of morphologies that are naturally produced by our simulations. The first and second rows display relatively isolated, compact systems, though the central source in the second row is only marginally detected in the rest-frame UV due to its relatively low UV flux ($M_{\rm UV } \simeq -19.6$) compared to the first case ($M_{\rm UV } \simeq -21.4$). The third row illustrates a partially resolved host galaxy surrounding the central source. The fourth row shows a nearby companion suggesting an ongoing merger, while the fifth row exhibits prominent off-centre stellar clumps. 

%Interestingly, most of the morphological diversity is confined to the short-wavelength images. Toward longer wavelengths, the images become remarkably homogeneous, with all systems appearing as compact, unresolved nuclei as the dust-attenuated AGN progressively dominates the observed emission.
Interestingly, most of the morphological diversity is confined to the short-wavelength images. Toward longer wavelengths, the images become remarkably homogeneous, with all systems appearing as compact, unresolved nuclei. Although the broader NIRCam PSF at longer wavelengths contributes to this apparent rest-frame optical compactness, the trend is primarily physical: as the effects of dust attenuation diminish toward longer wavelengths, emission from the obscured AGN increasingly outshines that of the host galaxy. Conversely, at $z=6$--$9$, the NIRCam SW channels probe the rest-frame UV, where host-galaxy structures remain more readily detectable. The NIRCam SW bands therefore provide the clearest observational window for revealing the morphological diversity of the LRD population.

Both the broad morphological complexity in the rest-frame UV and the pronounced increase in brightness from the rest-UV to the rest-optical reproduced by our simulations closely resemble \emph{JWST}/NIRCam observations of LRDs,  where clumpy, interacting, and extended rest-frame UV structures frequently coexist with compact unresolved nuclei dominating the rest-frame optical emission \citep[e.g.,][]{Baggen_2023, baggen_small_2024, baggen2025resolvingcomplexmultiscalemorphology, Baggen_2026, labbe_unambiguous_2024, Tanaka_2024, Chen_offcenter_2025, Merida_2025,  Rinaldi_2025, rinaldi2026dotlrdlikenucleusheart, DEugenio_2026, Torralba_2026, yanagisawa2026venusfaintlittlered}. 

\vspace{-0.2cm}
\subsubsection{UV Brightness and Optical Dust Attenuation} \label{sec:uv}
\begin{figure*}
    \centering
        \includegraphics[width = \linewidth]{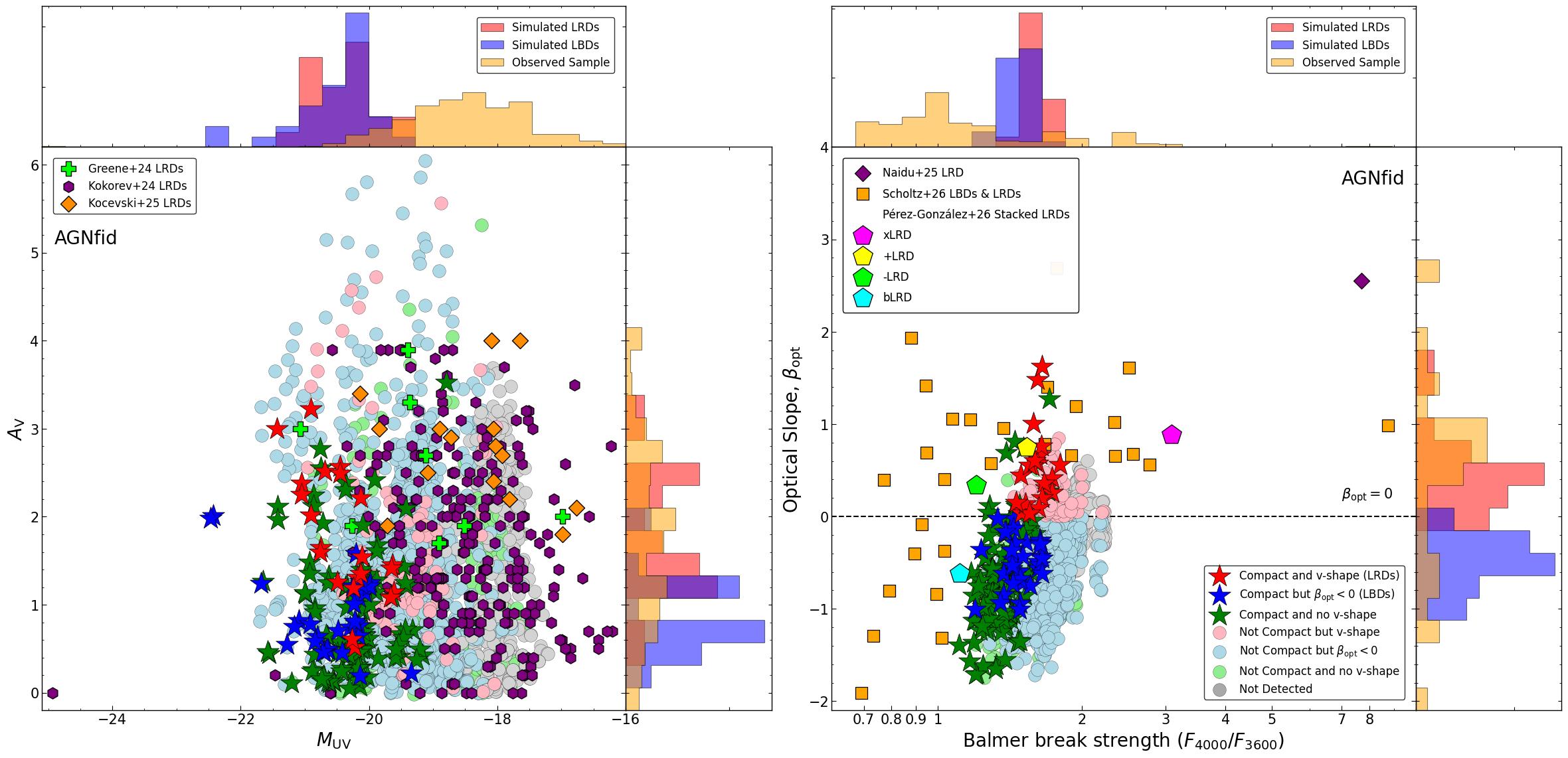}
        \caption{Observed properties of our \emph{AGNfid} sample. Star markers denote sources that are compact ($\kappa < 1.7$), whereas circles indicate extended sources. Red (pink) markers designate sources satisfying all LRD photometric selection criteria (simulated LRDs), blue (light-blue) markers signify sources that violate only the $\beta_{\rm opt}$ cutoff (simulated LBDs), while green (light-green) markers identify sources that fail the rest-UV selection ($\beta_{\rm UV}$) for compact (extended) populations. Gray markers correspond to sources falling below the adopted detection threshold (Sec.~\ref{res:LRDphoto}). Marginal histograms along the top and right axes show the normalised probability distributions for simulated LRDs (red), simulated LBDs (blue), and literature observed samples (orange). {\bf Left panel:} Visual attenuation ($A_{\rm V}$) versus rest-frame absolute UV magnitude ($M_{\rm UV}$). For empirical comparison, we overplot observed LRD populations from \citealt{greene_uncover_2024} (green pluses), \citealt{kokorev_2024} (purple hexagons), and \citealt{kocevski_rise_2024} (orange diamonds). {\bf Right panel:} Rest-frame optical spectral index ($\beta_{\rm opt}$) as a function of the Balmer break strength ($F_{4000}/F_{3600}$). For comparison, we overplot the MoM-BH* LRD from \citealt{naidu2025blackholestarreveals} (purple diamond), the empirical sample of observed LRDs, LBDs, and X-ray AGN from \citealt{scholtz2026littleredbluedots} (orange squares), and values derived using stacked LRD spectra (xLRDs in magenta, $+$LRDs in yellow, $-$LRDs in lime,  bLRDs in cyan) from \citealt{perezgonzalez2026littlereddotsphotometric} (coloured pentagons). The horizontal dashed line marks a flat optical continuum slope ($\beta_{\rm opt} = 0$).
        } %\SC{Should I overplot Akins, Furtak, and other LRDs? Which A_V to use?}
        \label{fig_obs:5}
\end{figure*}
The left panel of Fig.~\ref{fig_obs:5} compares the rest-frame UV absolute magnitude ($M_{\rm UV}$) and visual attenuation ($A_{\rm V}$) of our simulated \emph{AGNfid} sources with observed LRD samples \citep{greene_uncover_2024,kokorev_2024, kocevski_rise_2024}. The observational $A_{\rm V}$ values are model-dependent, as they are inferred from SED fitting under different assumptions regarding the relative contributions of the AGN and host galaxy\footnote{In \citetalias{kocevski_rise_2024}, $A_{\text{V}}$ was estimated using \texttt{CIGALE}~\citep{Boquien_2020, yang2020, Yang_2022} assuming a composite system where a dust-obscured AGN (parametrised via the \texttt{SKIRTOR} library; \citealt{Stalevski_2012, Stalevski_2016}) dominates the rest-frame optical, whereas unattenuated stellar light \citep{Bruzual2003} leaks into the rest-frame UV. \citealt{greene_uncover_2024} derived $A_{\text{V}}$ by fitting an SDSS composite AGN template \citep{Vanden_2001}, assuming the rest-frame optical is fully AGN-dominated. Similarly, \citealt{kokorev_2024} performed SED fits using a two-component, AGN-only model combining the SDSS quasar template with near-infrared spectra \citep{Glikman_2006}, assuming a negligible host-galaxy contribution in the optical and treating the rest-frame UV as scattered AGN light.}. Nevertheless, all of these analyses assume that the rest-frame optical emission is dominated by an obscured AGN, an assumption that closely matches the physical picture emerging from our simulations (see Sec.~\ref{res:interpretation}), making the comparison broadly self-consistent.

Our simulated sample occupies a region of UV luminosities ($-22.5<M_{\rm UV}<-17$) and visual attenuation ($0.5<A_{\rm V}<6$) in good agreement with the observed population. In particular, our compact sources, including the simulated LRD candidates, preferentially populate the brightest ($-22.5<M_{\rm UV}<-19.5$) and moderately obscured ($A_{\rm V}\lesssim3.5$) part of the distribution. This behaviour suggests that the observed LRD colours are preferentially reproduced by intrinsically luminous systems (see Sec.~\ref{res:accretion}) viewed through lines of sight that provide substantial, but not complete, attenuation of the central AGN.

\vspace{-0.3cm}
\subsubsection{Balmer Break} \label{res:balmerbreak}
% Should we include more data that is not only from Scholtz since that only focusses on the cases with strong broad Halpha emission. 
In the right panel of Fig.~\ref{fig_obs:5}, we map the Balmer break strength of our simulated \emph{AGNfid} objects against their rest-frame optical spectral index, $\beta_{\rm opt}$. For comparison, we overplot the sample of observed Little Red Dots (LRDs) and Little Blue Dots (LBDs), the MOM-BH*1 LRD from \citealt{naidu2025blackholestarreveals}, and the Balmer break strengths derived from empirical stacked spectra \citep{perezgonzalez2026littlereddotsphotometric}. These stacks are categorised by optical-to-UV luminosity ratio ($L_{1500}/L_{2500}$) into four subsamples: extreme (xLRDs), red ($+$LRDs), blue ($-$LRDs), and broad-line blue (bLRDs).
Following the observational definition, the Balmer break strength is quantified as the median flux ratio across the discontinuity, $F_{4000}/F_{3600}$, where $F_{3600}$ and $F_{4000}$ are the fluxes at $3600~\angstrom$ and $4000~\angstrom$~, respectively. The majority of observed Balmer break strengths span $\sim 0.7$ to $3.0$, though extreme outliers exist, such as "Cliff" \citep{DeGraff2025} with $F_{4000}/F_{3600} \sim 9.0$ and MoM-BH* \citep{naidu2025blackholestarreveals} with $F_{4000}/F_{3600} \sim 7.7$. Our simulated \emph{AGNfid} sources are characterised by Balmer break strengths in the range $\sim 1-2$. 

\begin{figure}
    \centering
        \includegraphics[width = \linewidth]{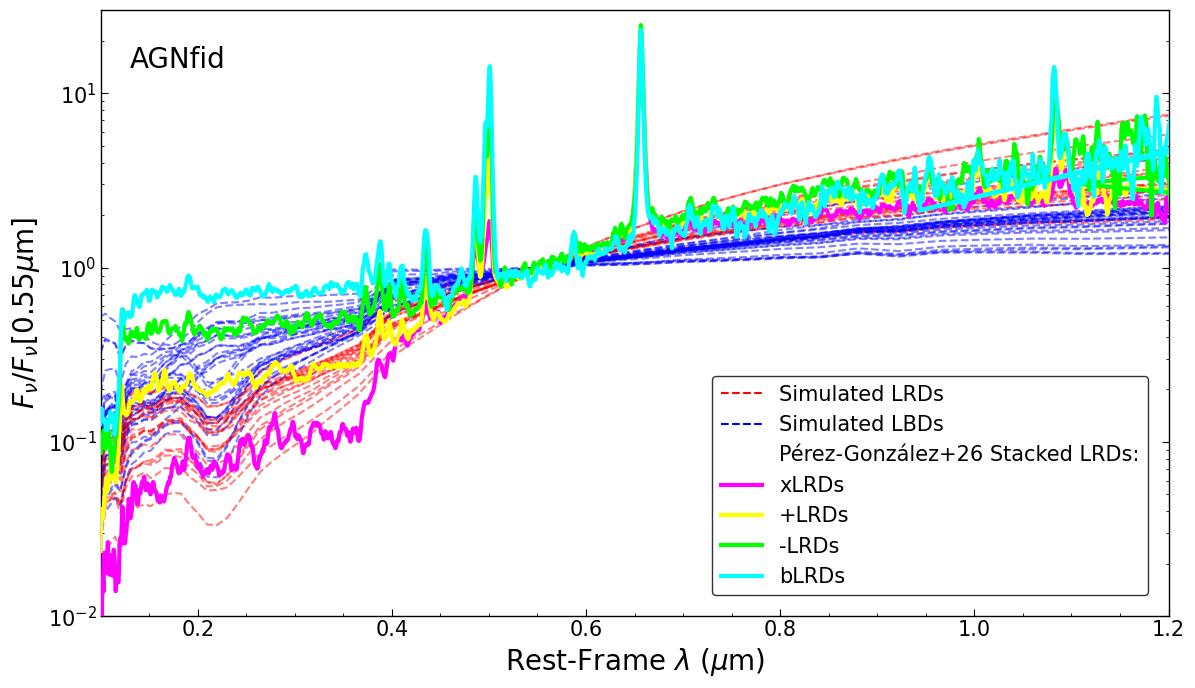}
        \caption{Comparison of rest-frame SED shape for our simulated \emph{AGNfid} sample against observational composite spectra from \citealt{perezgonzalez2026littlereddotsphotometric}. Red (blue) dashed curves represent normalised SEDs for our simulated compact LRD (LBD) candidates, defined by their compactness and v-shape (blue optical slope, $\beta_{\rm opt} < 0$) criteria. For comparison, solid curves depict the empirical stacked spectra of xLRDs (magenta), $+$LRDs (yellow), $-$LRDs (lime), and bLRDs (cyan). All spectra are normalised to flux density at rest-frame $\lambda = 0.55\,\mu{\rm m}$.}
        \label{fig_obs:spec_comp}
\end{figure}
\begin{figure*}[h]
    \centering
        \includegraphics[width = \linewidth]{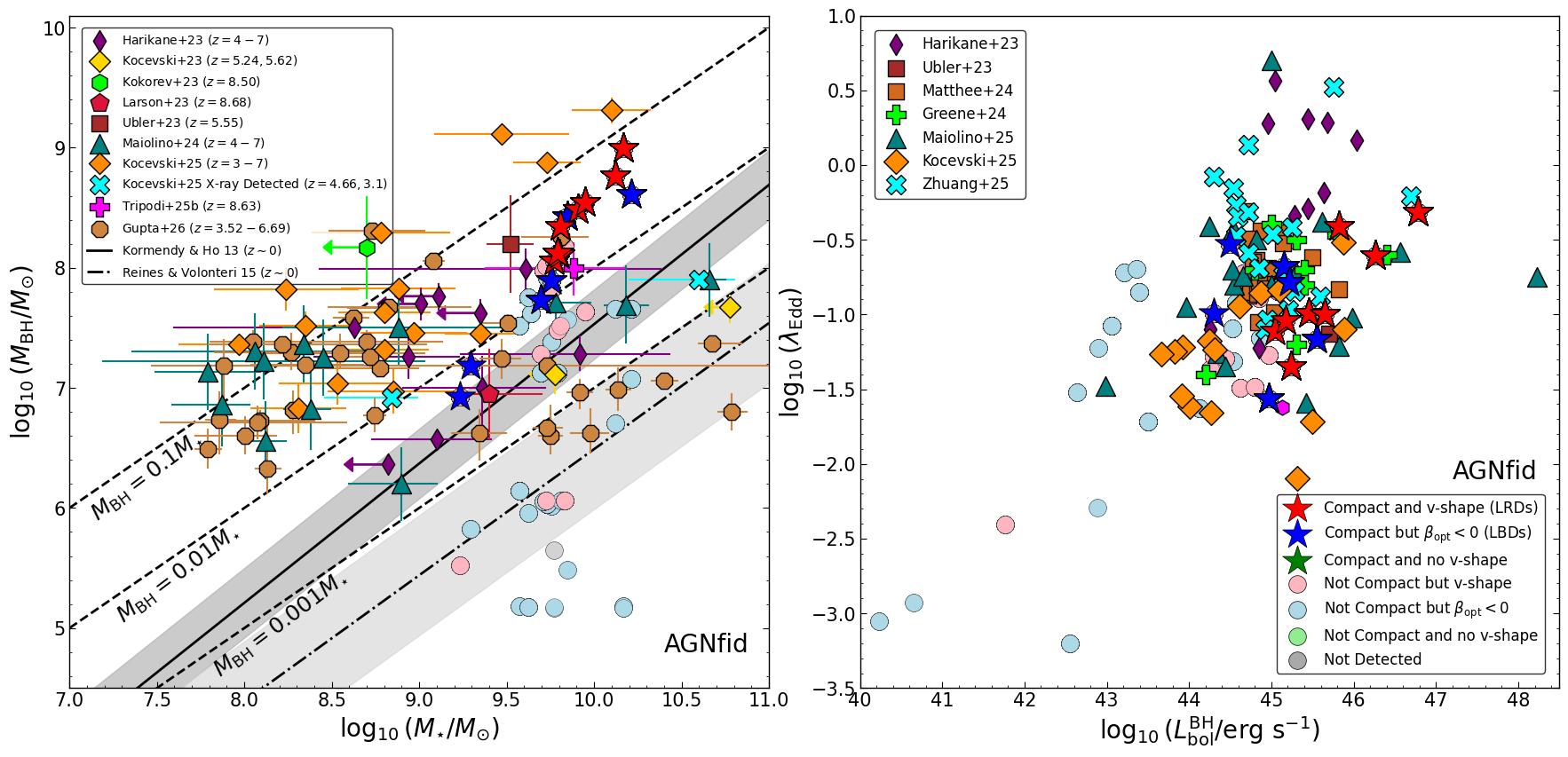}
        \caption{Intrinsic physical properties of the simulated sources in the \emph{AGNfid} suite. Star markers denote sources that are compact, whereas circles indicate extended sources. Red (pink) markers designate sources satisfying all LRD photometric selection criteria, blue (light-blue) markers signify sources that violate only the $\beta_{\rm opt}$ cutoff, while green (light-green) markers identify sources that fail the rest-UV selection ($\beta_{\rm UV}$) for compact (extended) populations. Gray markers correspond to sources falling below the adopted detection threshold. {\bf Left panel:} Black hole mass ($M_{\rm BH}$) versus host galaxy stellar mass ($M_{\star}$) relation for our simulated sample. For empirical comparison, we overplot high-redshift LRDs and spectroscopically confirmed broad-line AGN from recent \emph{JWST} literature \citep{harikane2023, kocevski_hidden_2023, kocevski_rise_2024, Kokorev_2023, larson2023, ubler_ga-nifs_2023, Maiolino_2024, Tripodi_2025, gupta2026rapidevolutionobservedmbhm}. The local $z \sim 0$ scaling relations from \citealt{Kormendy13} (solid black line) and \citealt{reines2015} (dash-dotted black line) are shown as benchmarks. Diagonal dashed gray lines indicate constant black hole-to-stellar mass ratios of $M_{\rm BH}/M_{\star} = 0.1$, $0.01$, and $10^{-3}$. {\bf Right panel:} Eddington ratio ($\lambda_{\rm Edd}$) versus black hole bolometric luminosity ($L_{\rm bol}^{\rm BH}$) for our simulated sample. Observed high-redshift LRDs with spectroscopic broad-line confirmation \citep{matthee_little_2024, greene_uncover_2024, kocevski_rise_2024, Juodbalis_2026, zhuang2025nexusspectroscopiccensusbroadline} and Type~I AGN from various studies \citep{harikane2023, ubler_ga-nifs_2023, maiolino_jwst_2025} are overplotted for comparison.}
        \label{fig_obs:8}
\end{figure*}
Although our simulated compact systems reproduce the Balmer break strengths exhibited by approximately $40\text{--}50\%$ of the observed LRD and LBD population, they do not reproduce systems with $F_{4000}/F_{3600} < 1$ or $F_{4000}/F_{3600} \gtrsim 2$. The former may arise from nebular continuum emission, which is not included in our RT calculations. For what concerns the latter, two regimes must be considered separately (see Fig. A.1 in \citealt{Wang_2024}): i) $2\lesssim F_{4000}/F_{3600} \lesssim 2.5$. These values can be reached for evolved stellar populations with ages $t_{\star} \simeq 0.3\text{-}1.0\,\mathrm{Gyr}$; ii) $F_{4000}/F_{3600} \gtrsim 2.5$. These extreme values cannot have a pure stellar origin and are instead attributed to absorption by extremely high neutral hydrogen column densities ($N_{\rm HI} \sim 10^{24-26} \rm cm^{-2}$) along the line of sight \citep{DeGraff2025, naidu2025blackholestarreveals, ji_blackthunder_2025, inayoshi_extremely_2025}. 

For what concerns our LRD candidates, the stellar populations of compact \emph{AGNfid} objects are predominantly young, with mass-weighted ages of $t_{\star} \simeq 0.1\text{--}0.2\,\mathrm{Gyr}$. Therefore, these are not evolved enough to explain intermediate Balmer break strengths ($2\lesssim F_{4000}/F_{3600} \lesssim 2.5$). Furthermore, our simulations span lower line-of-sight column densities ($N_{\rm H} \simeq 3 \times 10^{22}\text{--}10^{23}\,\mathrm{cm^{-2}}$) and do not explicitly model excited $n=2$ hydrogen absorption physics required to reproduce such pronounced Balmer breaks.

Interestingly, our simulated LBDs cluster tightly at $\beta_{\rm opt}<0$ while exhibiting Balmer break strengths comparable to those of our LRD candidates. Combined with the simulation results, which show that the two populations host the same underlying stellar populations, this suggests that the observed differences in optical colour are primarily driven by the presence of substantial dust along the line of sight in LRDs.

To assess how these Balmer break strengths translate into overall continuum shapes, Fig.~\ref{fig_obs:spec_comp} compares the rest-frame SEDs of our simulated LRDs and LBDs against the \citet{perezgonzalez2026littlereddotsphotometric} empirical stacks. The simulated LRD SEDs broadly reproduce the rest-frame UV and optical continua of the xLRD ($L_{1500}/L_{2500}>6.3$) and $+$LRD ($3.1<L_{1500}/L_{2500}<6.3$) stacks, with the notable exception of the pronounced Balmer break exhibited by the xLRDs. By contrast, simulated LBDs more closely resemble the bluer $-$LRD ($1.8<L_{1500}/L_{2500}<3.1$) and bLRD ($L_{1500}/L_{2500}<1.8$) stacks in the rest-frame UV, though their continua become shallower at $\lambda_{\rm rest}\gtrsim0.55\,\mu\mathrm{m}$. Consistent with the right panel of Fig.~\ref{fig_obs:5}, the Balmer break strengths predicted for both simulated populations overlap primarily with those of the $+$LRD, $-$LRD, and bLRD stacks. The substantially stronger breaks observed in the xLRD stack are not reproduced by our simulations and likely probe a different physical regime, involving either more evolved stellar populations or considerably higher line-of-sight column densities, as discussed above.
\vspace{-0.3cm}

\subsection{Intrinsic Physical Properties} \label{res:properties}
In this section, we examine the intrinsic physical properties of the synthetic systems selected as LRD candidates and compare them with inferred values from observations. %First, we evaluate where our simulated population lies on the black hole mass–stellar mass ($M_{\rm BH}$–$M_{\star}$) scaling relation in Section \ref{res:bhstmass}. Then, in Section \ref{res:accretion}, we analyze the central black hole Eddington ratios to determine how their growth profiles compare with current observational estimates.

\subsubsection{The Black Hole–Stellar Mass Relation} \label{res:bhstmass}
The left panel of Fig. \ref{fig_obs:8} compares the $M_{\rm BH}$--$M_{\star}$ relation\footnote{We define the stellar mass as the total mass of star particles enclosed within $0.1\,r_{\rm vir}$ of the central SMBH, where $r_{\rm vir}$ is the virial radius of the host halo at the corresponding simulation snapshot \citep{Zana_2022}.} for our simulated \emph{AGNfid} sample with observations of spectroscopically confirmed Type I AGN \citep[e.g.,][]{harikane2023, kocevski_hidden_2023, kocevski_rise_2024, Kokorev_2023, larson2023, ubler_ga-nifs_2023, Maiolino_2024}. The simulated galaxies span stellar masses of $\log(M_{\star}/M_{\odot}) \sim 8.9\text{--}9.8$ and black hole masses of $\log(M_{\rm BH}/M_{\odot}) \sim 6.9\text{--}9.0$. The LRD selection isolates a population occupying the upper end of both the stellar and black hole mass distributions, with $\log(M_{\star}/M_{\odot}) \sim 9.8\text{--}10.2$ and $\log(M_{\rm BH}/M_{\odot}) \sim 8.0\text{--}9.0$.

%For comparison, we overplot a compilation of high-redshift LRDs and broad-line AGN with robust broad emission-line detections from recent JWST surveys   %We additionally compare our simulated population with the local $z\sim0$ $M_{\rm BH}$--$M_\star$ scaling relations for active galaxies \citep{reines2015} and early-type bulges \citep{Kormendy13}.

In agreement with data, our LRD candidates are systematically overmassive relative to the local scaling relations. This offset is expected for two reasons. First, by targeting the central galaxy of the most massive halo in our simulation volume, our sample is intrinsically biased toward the upper end of both the stellar and BH mass distributions. Second, the fiducial simulation adopts an AGN feedback efficiency calibrated to reproduce the observed normalisation of the $M_{\rm BH}$--$M_{\star}$ relation at $z\sim6$ for luminous quasars \citep[e.g.,][]{Wang_2010, pensabene2020}.

We identify a subtle trend across the $M_{\rm BH}$--$M_{\star}$ plane: systems with higher $M_{\rm BH}/M_{\star}$ ratios are more likely to satisfy our photometric LRD selection (red stars), whereas lower mass ratios are more commonly associated with unselected compact systems (blue stars). This trend may arise because more massive black holes produce a stronger optical AGN continuum, making the characteristic LRD colour selection easier to satisfy; we discuss this interpretation in Sec. \ref{res:interpretation}. Notably, several compact \emph{AGNfid} sources that lack a "V-shape" occupy a lower envelope that lies much closer to the local scaling relations. This is consistent with recent observations of LBDs, which appear to be significantly better in agreement with the local $M_{\rm BH}$--$M_{\star}$ relation than their red counterparts \citep{madau_maioliono2026}. %Nevertheless, the final broadband colours are determined not only by the intrinsic black hole and stellar masses, but also by the local dust column density, line-of-sight geometry, and the adopted dust grain model, leading to substantial overlap between selected and unselected systems at intermediate masses.

%Should we mention about possibly these values are overestimated for Bhs? Cause it goes against our cases? JudzBalis 2026 for writeup
% Consider plotting only for a partricular dust type, so we can see this trend clearly, cause dust also influence the selection.
% Please consider if the stellar mass is properly calculated since does it calculate only a part of the stellar mass? Also, should only the LoS value should be considered?

\subsubsection{Bolometric Luminosity and Eddington Ratio} \label{res:accretion}

% Have to plot more objects:
The right panel of Fig. \ref{fig_obs:8} compares the predicted distribution of Eddington ratios ($\lambda_{\rm Edd} \equiv L_{\rm bol}/L_{\rm Edd}$) as a function of the BH bolometric luminosity with that inferred for observed high-redshift LRDs \citep{matthee_little_2024, greene_uncover_2024, kocevski_rise_2024, zhuang2025nexusspectroscopiccensusbroadline, Tripodi_2025} and Type I AGN \citep{harikane2023, ubler_ga-nifs_2023, maiolino_jwst_2025, gupta2026rapidevolutionobservedmbhm}.\footnote{Bolometric luminosities are estimated from the extinction-corrected broad H$\alpha$ luminosity using the scaling relation $L_{\rm bol}^{\rm BH} \simeq 130\,L_{\rm H\alpha}^{\rm broad}$ \citep{Stern&Laor}, while black hole masses are derived from single-epoch virial estimators.} For our simulated LRD candidates, $L_{\rm bol}$ is derived directly from the instantaneous black hole accretion rates provided by the hydrodynamic simulations.  Our sample reproduces the bulk of the observed population, spanning $0.05 < \lambda_{\rm Edd} < 0.5$, whereas high-redshift AGN displays both sub- and super-Eddington accretion states. In our hydrodynamic framework \citepalias{Valentini:2021}, gas accretion onto the central black hole is capped at the Eddington limit and self-regulated by isotropic thermal AGN feedback, preventing the simulated systems from entering super-Eddington phases.

However, this comparison should be interpreted with caution, as observationally inferred Eddington ratios are subject to substantial systematic uncertainties in both BH mass\footnote{Despite these systematic uncertainties, the few available dynamical constraints are broadly consistent with single-epoch virial estimates. In particular, this agreement has been demonstrated for the prototypical LRD Abell2744-QSO1 \citep{Juodbalis_2026}.} and bolometric luminosity estimates. Recent studies have argued that the BLRs of JWST AGN may differ systematically from those of local AGN, potentially exhibiting substantially larger covering factors than assumed in standard local calibrations \citep[e.g.,][]{maiolino_jwst_2025}. A larger BLR covering factor would enhance the broad H$\alpha$ luminosity at fixed bolometric luminosity and could therefore lead to Eddington ratios being overestimated by factors of $\sim 1.4$--$3$. Moreover, \citet{Greene2026} showed that standard bolometric corrections may overestimate $L_{\rm bol}$ by up to an order of magnitude if the red optical continuum traces intrinsic emission from the central engine rather than strong dust attenuation. These effects may alleviate the apparent tension between our simulated population, which spans $0.05 < \lambda_{\rm Edd} < 0.5$, and the handful of observed sources inferred to accrete at super-Eddington rates. Nevertheless, because our accretion prescription is capped at the Eddington limit, genuinely super-Eddington phases cannot be captured by the present simulations.

\begin{figure}[h]
    \centering
        \includegraphics[width = \linewidth]{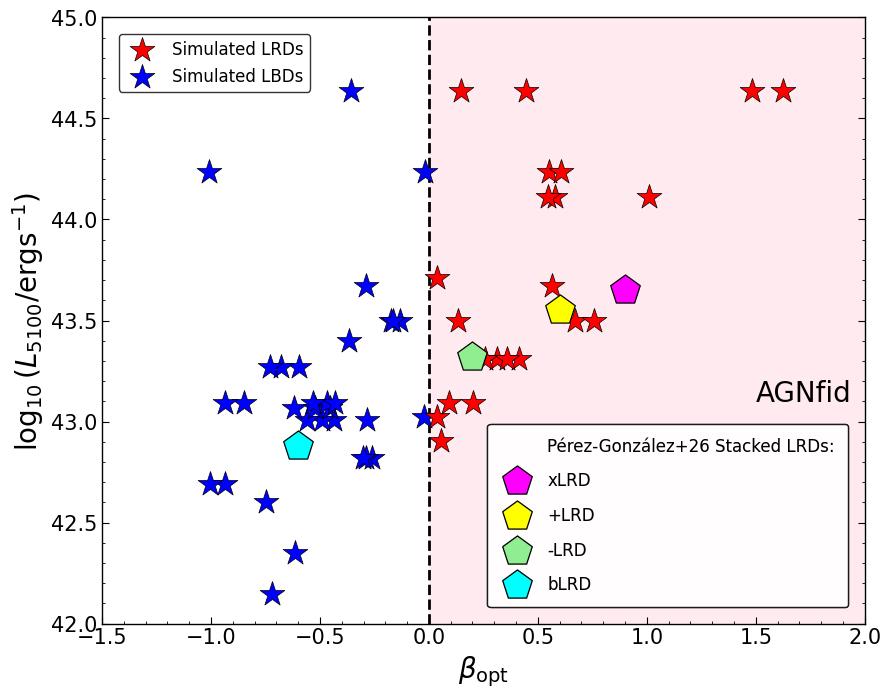}
        \caption{Optical continuum slope ($\beta_{\rm opt}$) versus monochromatic luminosity at rest-frame $5100\,\text{\AA}$ ($\log_{10}( L_{5100}/\rm erg~s^{-1})$) for our simulated LRDs and LBDs. Red star markers denote simulated LRDs satisfying all LRD photometric criteria, whereas blue star markers designate simulated LBDs, which violate solely the optical slope selection ($\beta_{\rm opt} < 0$). For comparison, coloured pentagons indicate median values inferred from the stacked NIRSpec/PRISM spectra of four LRD subtypes (xLRD in magenta, $+$LRD in yellow, $-$LRD in green, and bLRD in cyan) from \citealt{perezgonzalez2026littlereddotsphotometric}. The pink shaded region ($\beta_{\rm opt} > 0$) demarcates the photometric boundary introduced by \citetalias{kocevski_rise_2024} which separates classical LRDs from their bluer counterparts.}
        \label{fig_obs:l5100}
\end{figure}

Given these uncertainties, we further explore the relation between the directly observable rest-frame $5100\,\text{\AA}$ monochromatic continuum luminosity ($L_{5100}$) and the optical continuum slope ($\beta_{\rm opt}$). This comparison provides a complementary view that does not require estimates of either $M_{\rm BH}$ or $L_{\rm bol}$ \citep{rinaldi2026waytallytaleimpact, barrufet2026orientationevidencedistinctphysical}. In Fig.~\ref{fig_obs:l5100}, we plot $\log_{10}(L_{5100})$ against $\beta_{\rm opt}$ for our simulated LRDs and LBDs. Our modeled systems exhibit a clear positive correlation between $L_{5100}$ and $\beta_{\rm opt}$, in strong agreement with the trends inferred from the stacked LRD spectra \citep[xLRDs, $+$LRDs, $-$LRDs, and bLRDs;][]{perezgonzalez2026littlereddotsphotometric}. This continuous trend supports recent frameworks proposing that LRDs and LBDs belong to a single underlying physical population, separated primarily by an arbitrary threshold in $\beta_{\rm opt}$, with variations in optical reddening driven by intrinsic AGN emission \citep{rinaldi2026waytallytaleimpact} or dust obscuration \citep{scholtz2026littleredbluedots, madau_maioliono2026}, as detailed in Sec.~\ref{res:interpretation}.

\subsection{Physical origin of the simulated LRDs} \label{res:interpretation}
The previous sections have shown that our simulated LRD candidates reproduce a wide range of the observed properties of the LRD population. They naturally satisfy the characteristic V-shaped continuum and compact morphology, and are qualitatively consistent with the observed JWST imaging. %, reproduce the flat rest-frame mid-infrared continuum, and remain compatible with current far-infrared upper limits. 
In addition, both their inferred observational properties ($M_{\rm UV}$, $A_V$) and intrinsic physical properties ($M_{\rm BH}$, $M_{\star}$, and $\lambda_{\rm Edd}$) broadly agree with those inferred for observed systems. 

\begin{figure}[h]
    \centering
    \includegraphics[width = \linewidth]{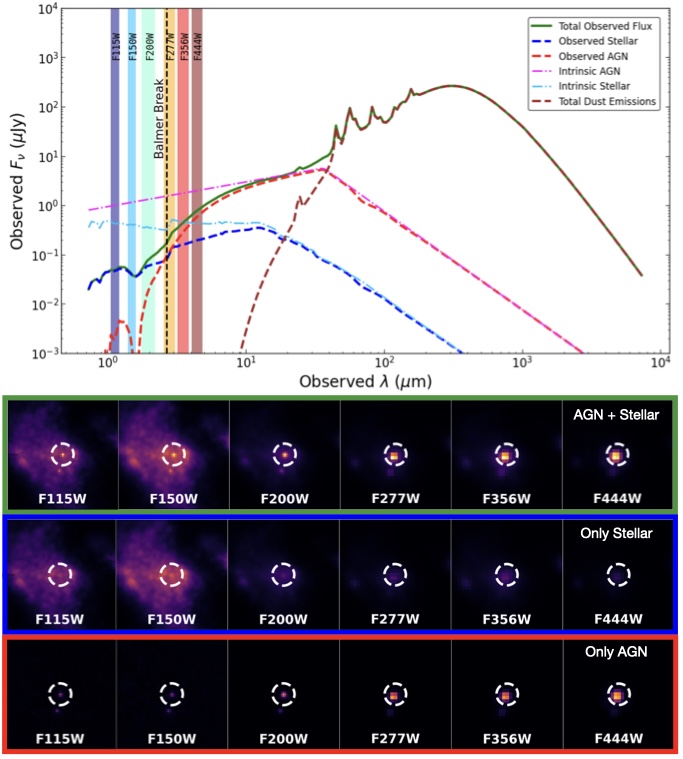}
    \caption{Spectral energy distribution (SED) and spatial decomposition of a representative simulated LRD candidate at $z=6.3$, post-processed using the \citetalias{weingartner2001} Milky Way dust model. The system hosts a central massive black hole ($M_{\rm BH}\simeq3.5\times10^8\,M_{\odot}$), a host galaxy stellar mass of $M_{\star}\simeq2.9\times10^9\,M_{\odot}$, and a total dust mass of $M_{\rm d}\simeq5.7\times10^5\,M_{\odot}$. \textbf{Top panel:} Rest-UV to far-infrared SED, where the solid green curve represents the total emergent flux. Dashed curves show individual dust-attenuated observed components: stars (blue), AGN (red), and thermal dust re-emission (brown). Dash-dotted profiles represent unattenuated, intrinsic emission from stars (cyan) and the central engine (magenta). Shaded vertical bands designate \emph{JWST}/NIRCam filter coverages utilised in our photometric selection criteria (Sec.~\ref{sec:SEDs}), with the rest-frame Balmer break marked by the vertical dashed black line. \textbf{Bottom panel:} Corresponding noiseless NIRCam image stamps ($0.2''$ diameter apertures shown as white dashed circles) illustrating the spatial decomposition across filter bands for total observed emission (green border; top row), stellar emission (blue border; middle row), and central AGN emission (red border; bottom row).}
    \label{fig_obs:9b}
\end{figure}

To identify the physical mechanisms responsible for the observed LRD phenomenology, in Fig.~\ref{fig_obs:9b} we examine the spectral decomposition of the SED of a simulated LRD candidate at $z=6.3$. To isolate the stellar and AGN contributions, we performed additional radiative transfer calculations with the AGN emission switched off. Subtracting this stellar-only spectrum from the full radiative transfer solution directly yields the contribution of the dust-attenuated AGN continuum.\footnote{Primary unscattered photons follow linear attenuation through a fixed dust grid, making this subtraction an exact separation for the direct continuum. Secondary scattering processes and Monte Carlo sampling can introduce minor non-linearities between separate runs, but these remain second-order effects across our rest-frame optical analysis.}

As shown in Fig.~\ref{fig_obs:9b}, the observed rest-frame UV continuum is dominated by the host galaxy stellar population (dashed blue line), whereas the AGN emission is heavily attenuated by the dense central dust column. This difference reflects the distinct spatial distributions of the two emitting components: the stellar population extends over kiloparsec scales, while the AGN resides at the centre of the galaxy, where the gas and dust densities naturally peak, leading to much stronger attenuation of the nuclear continuum. The resulting stellar continuum remains sufficiently blue to satisfy the $\beta_{\rm UV}< -0.37$ selection criterion. Redward of the Balmer break, the dust-attenuated AGN continuum (dashed red line) emerges and rapidly overtakes the stellar emission, producing the steep optical rise required to satisfy the $\beta_{\rm opt}>0$ criterion. The stellar Balmer break naturally falls close to the JWST/NIRCam F277W bandpass at these redshifts, but its apparent strength is substantially diluted by the emerging AGN continuum. Consequently, signatures of intrinsic AGN variability and duty cycle will primarily manifest in the rest-frame optical, rather than in the rest-frame UV, which remains anchored by the comparatively stable stellar host. The resulting variations in AGN luminosity can alter the degree of Balmer-break dilution and the optical slope ($\beta_{\rm opt}$), potentially shifting sources into or out of the LRD colour selection as their black holes evolve through different accretion phases.%\SC{Add a comment about the AGN variability and duty cycle affecting the rest-optical part of the LRD spectrum.}

\begin{figure}
    \centering
    \includegraphics[width = \linewidth]{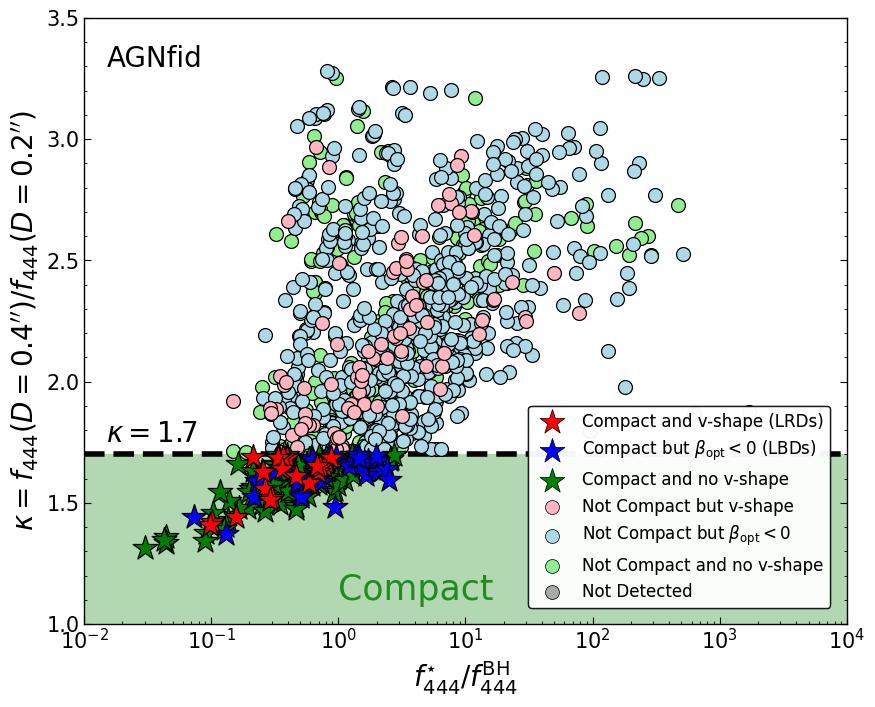}
    \caption{Compactness ratio ($\kappa$) as a function of the observed stellar-to-black hole flux ratio ($f^{\star}_{\text{444}}/f^{\rm \rm BH}_{\text{444}}$) in the JWST/NIRCam F444W bandpass. The dashed horizontal line denotes the threshold ($\kappa = 1.7$) separating extended systems from the compact region (shaded green area). Red (pink) markers designate sources satisfying all LRD photometric selection criteria, blue (light-blue) markers signify sources that violate only the $\beta_{\rm opt}$ cutoff, while green (light-green) markers identify sources that fail the rest-UV selection ($\beta_{\rm UV}$) for compact (extended) populations. Gray markers correspond to sources falling below the adopted detection threshold.}
    \label{fig_obs:9a}
\end{figure}

If the rest-frame optical continuum is dominated by the unresolved AGN, one naturally expects the source to appear morphologically compact. To investigate the origin of the compact morphology, Fig.~\ref{fig_obs:9a} compares the compactness parameter, $\kappa$, with the dust-attenuated stellar-to-AGN flux ratio, $f_{444}^{\star}/f_{444}^{\rm AGN}$, measured within the same $0.4''$ aperture. A clear monotonic trend emerges: as the stellar contribution to the observed F444W flux increases, the sources become progressively less compact (larger $\kappa$). Conversely, the most compact systems are those in which the unresolved AGN continuum dominates the rest-frame optical emission.

The scatter around this relation reflects differences in the intrinsic sizes of the host galaxies. At fixed stellar-to-AGN flux ratio, more extended stellar distributions produce larger values of $\kappa$. Nevertheless, Fig.~\ref{fig_obs:9a} clearly shows that an AGN-dominated rest-frame optical continuum is a necessary condition for a source to appear as a compact LRD. In other words, the observed compactness is primarily controlled by whether the AGN dominates the rest-frame optical continuum, while the intrinsic size of the host galaxy introduces only a secondary modulation.

Finally, the same physical picture naturally explains the different behaviour of the MW and SMC dust models. SMC dust configurations produce a larger fraction of compact sources because the SMC extinction curve yields a weaker attenuation at rest-frame optical wavelengths than the MW model of \citetalias{weingartner2001}, allowing the AGN continuum to dominate the observed F444W emission more readily. However, because the SMC extinction curve lacks the prominent $2175\,\AA$ absorption feature and exhibits a steeper UV extinction law than the MW model, it reddens the rest-frame UV continuum more efficiently, preventing many of these systems from satisfying the LRD colour selection.

Taken together, Fig.~\ref{fig_obs:9a} and Fig.~\ref{fig_obs:9b} demonstrate that the defining observational properties of LRDs arise naturally from the interplay between stellar emission, a dust-attenuated AGN continuum, and the line-of-sight dust geometry. In our simulations, LRDs therefore do not emerge as a fundamentally distinct class of galaxies, but rather as ordinary high-redshift AGN-host galaxies viewed through a particular dust geometry.

\section{Limitations and future prospects}\label{discussion}
Our simulations demonstrate that many of the defining observational properties of LRDs arise naturally within a standard AGN--host galaxy scenario. The remaining discrepancies therefore provide a useful guide to the additional physical ingredients that may be required to reproduce the full diversity of the observed population. In this section, we discuss the main limitations of our current modelling and outline the physical ingredients required for the next generation of cosmological simulations.

\subsection{The missing dusty torus} 
One limitation of our radiative transfer calculations is that they do not include absorption and re-emission by a dusty torus. This is particularly relevant because one of the most debated observational properties of LRDs is the apparent weakness or absence of the hot-dust emission typically associated with a parsec-scale torus. Photometric observations of individual LRDs often indicate an approximately flat continuum from the rest-frame optical to the NIR ($\lambda \simeq 1$-$3~\mu$m), with no clear hot-dust excess. This spectral shape has been interpreted as evidence for either the absence of an AGN or the lack of an associated hot-dust component \citep{Williams2023, Perez2024, Akins2024, Leung_2025, Wang_2025, Iani_2025, Setton_2025, Li_2025, Barro_2026}. 

However, this wavelength range is currently constrained mainly by relatively shallow photometry rather than by MIRI spectroscopy, and stacked analyses provide evidence that the average SED may not remain flat toward longer wavelengths \citep{Delvecchio_2025, perezgonzalez2026littlereddotsphotometric}. Moreover, an analysis of the wide-area SMILES MIRI survey showed that the vast majority of LRDs are not detected at MIRI wavelengths partly because most of them are already faint in NIRCam imaging \citep{rinaldi2026waytallytaleimpact}. Thus, unless an exceptionally large amount of dust reprocessing is assumed, their non-detection in current shallow, wide-field MIRI data can be a natural consequence of their intrinsic faintness and does not, by itself, demonstrate the absence of a torus.

Our simulated LRDs naturally reproduce the approximately flat continua inferred for many individual sources (see App. \ref{res:MIRI}); however, this agreement should be interpreted with caution. A dusty torus both obscures the central engine and reprocesses a substantial fraction of the AGN luminosity into the infrared. Including such a component could therefore alter both the predicted attenuation and the shape of the NIR/MIR SED. Consequently, our current models cannot establish whether the observed LRD population genuinely lacks a torus or whether approximately flat continua can arise despite its presence, particularly given the limited depth and wavelength coverage of the available observations.

However, the absence of a classical torus would not be unprecedented. A growing body of observational evidence indicates that hot-dust-deficient AGN exist across a broad range of redshifts and luminosities \citep[e.g.,][]{Hao_2010, Jiang_2010, Lyu_2017, Brown_2019, Son_2023}. In particular, the dusty torus may become weak or disappear in low-luminosity AGN if the accretion flow or disk wind can no longer sustain a geometrically and optically thick dusty structure. More generally, hot-dust deficiency has been attributed to intrinsic variations in torus structure \citep{Lyu_2022} or to the incomplete development of a stable dusty environment around rapidly growing BHs \citep[e.g.,][]{Cevenrino_2010, Hopkins_2024}. Future works should therefore incorporate physically motivated torus models and compare their predictions with deeper rest-frame NIR and MIR observations. This will be necessary to determine whether the observed diversity of LRD continua, from approximately flat SEDs to the reddest systems extending into the MIR \citep{Lyu_2024, Perez2024}, can be reproduced within a unified cosmological framework.

\subsection{X-ray weakness and super-Eddington accretion} 
An important observational property that our simulations fail to reproduce is the extreme X-ray weakness of LRDs. Applying standard bolometric corrections \citep{Duras2020} to our simulated LRD AGN in the luminosity range that matches the high-luminosity bin from \citealt{maiolino_jwst_2025} ($L_{\mathrm{bol}} \sim 10^{44.8}\text{--}10^{45.8}\,L_{\odot}$) yields the predicted hard X-ray fluxes of $F_{\mathrm{X}, 2\text{--}10\,\mathrm{keV}} \sim 10^{-16.2}\text{--}10^{-15.6}\,\mathrm{erg\,cm^{-2}\,s^{-1}}$. These values are approximately $20-60$ times higher than the stacked $90\%$ upper limit of $F_{\mathrm{X}, 2\text{--}10\,\mathrm{keV}} \lesssim 10^{-17.4}\,\mathrm{erg\,cm^{-2}\,s^{-1}}$ reported by \citet{maiolino_jwst_2025}, obtained assuming an X-ray spectral slope of $\Gamma=1.7$.
%Applying standard bolometric corrections \citep{Duras2020} to our simulated AGN yields intrinsic hard X-ray luminosities of $L_{\mathrm{X}, 2\text{--}10\,\mathrm{keV}} \sim 10^{44}\text{--}10^{45}\,\mathrm{erg\,s^{-1}}$, exceeding by several orders of magnitude the stacked upper limit of $L_{\mathrm{X}} \lesssim  10^{41.5}\,\mathrm{erg\,s^{-1}}$ reported by \citealt{maiolino_jwst_2025}. 
This discrepancy is not unique to our simulations, but reflects a broader observational puzzle: current X-ray observations indicate that LRDs, together with the wider population of UV-faint, spectroscopically confirmed broad-line AGN discovered by \textit{JWST}, are systematically X-ray weak \citep[e.g.,][]{ananna_x-ray_2024, yue_stacking_2024, maiolino_jwst_2025, tortosa2026xrayweaknesslittlered}. Two main physical explanations have been proposed to account for this X-ray weakness.

%The first invokes super-Eddington accretion, where photon trapping within slim accretion disks and/or dense radiation-driven outflows suppresses emergent X-ray emission from the nuclear corona \citep[e.g.,][]{madau_x-ray_2024, volonteri_exploring_2025, tortosa2026xrayweaknesslittlered,Madau_2026,madau_maioliono2026,Trinca2026}. In this context, super-Eddington AGN are expected to exhibit steeper X-ray spectra  \citep[up to $\Gamma\sim 4.5$; e.g.][]{Pacucci_2024} than their sub-Eddington counterparts ($\Gamma \sim 1.7$). This scenario is supported by recent observations of local AGN \citep{Tortosa_2022, Tortosa_2024} and $z\sim 6$ quasars \citep{Zappacosta_2023, Tortosa_2024_Hyperion, tortosa2026xrayweaknesslittlered}, although, at high-$z$, it is difficult to disentangle soft X-ray spectra from a cold corona.

The first invokes super-Eddington accretion, in which photon trapping within slim accretion disks and dense radiation-driven outflows suppresses the X-ray emission emerging from the nuclear corona \citep[e.g.,][]{madau_x-ray_2024,volonteri_exploring_2025,tortosa2026xrayweaknesslittlered,Madau_2026,madau_maioliono2026,Trinca2026}. In this scenario, super-Eddington AGN are expected to exhibit steeper X-ray spectra than their sub-Eddington counterparts ($\Gamma\sim1.7$), with photon indices reaching $\Gamma\sim 4.5$ \citep[e.g.,][]{Pacucci_2024}. This interpretation is consistent with recent observations of local AGN \citep{Tortosa_2022,Tortosa_2024} and $z\sim6$ quasars \citep{Zappacosta_2023,Tortosa_2024_Hyperion,tortosa2026xrayweaknesslittlered}. At high redshift, however, the limited quality and energy coverage of the available data make it difficult to distinguish an intrinsically steep power-law continuum from spectral curvature associated with a low coronal temperature.

%\textbf{ Theoretical models extend this value up to $\Gamma \sim 4.5$ \citep{Pacucci_2024}, although at high redshift such steep effective values cannot be disentangled from a low high-energy cutoff $E_{\mathrm{cut}}$.}

%Observationally, super-Eddington AGN systematically exhibit steeper X-ray spectra ($\Gamma \sim 2.0\text{--}2.5$; e.g., ) than their sub-Eddington counterparts ($\Gamma \sim 1.7$).

To estimate how a steep X-ray spectral slope affects the comparison between X-ray observations and our predictions, we first convert the observational upper limit reported by \citealt{maiolino_jwst_2025} using the NASA's \texttt{WebPIMMS} tool\footnote{\url{https://heasarc.gsfc.nasa.gov/cgi-bin/Tools/w3pimms/w3pimms.pl}}, and assuming $\Gamma=3.1$ \citep[i.e., the median value by][]{Pacucci_2024}. We obtain $F_{\mathrm{X}, 2\text{--}10\,\mathrm{keV}} \lesssim 10^{-16.6}\,\mathrm{erg\,cm^{-2}\,s^{-1}}$ (see more detailed calculations in Tortosa et al in prep.). %\SG{This must be removed, along with the Appendix. Furthermore, to compute the hard X-ray emission expected from our simulated sources, we proceed as in App. \ref{steep_Gamma}. We find that, assuming $\Gamma = 3.1$ reduces the predicted hard X-ray luminosities of our simulated sources to $F_{\mathrm{X}, 2\text{--}10\,\mathrm{keV}} \sim 10^{-17.0}\text{--}10^{-16.4}\,\mathrm{erg\,cm^{-2}\,s^{-1}}$,  which is consistent with the updated observational limit at the $\sim 0.8 - 3.3\,\sigma$ level.} and substitute with 
The revised limit is approximately six times less stringent than that obtained assuming $\Gamma=1.7$. Nevertheless, it remains a factor of $\sim 3-10$ below the X-ray fluxes expected from our simulated sources, indicating that a steep spectrum substantially alleviates, but does not completely remove, the tension with the current observational constraint. Fully reconciling the predicted fluxes with the observational limit would require $\Gamma \approx 5$, a photon index far steeper than typically observed.

The second explanation attributes the X-ray weakness to heavy, dust-free Compton-thick gas ($N_{\mathrm{H}} \gtrsim 10^{24}\,\mathrm{cm^{-2}}$) surrounding the broad-line region, which efficiently absorbs the X-ray emission while leaving the optical broad lines largely unaffected \citep[e.g.,][]{yue_stacking_2024, Juod_balis_2024, maiolino_jwst_2025}. To estimate whether such an obscuration can reconcile our simulated sources with the observed X-ray limits, we proceed as in App. \ref{compton_thick}. We find that our simulations predict column densities ($N_{\rm H} = 1.0^{+1.0}_{-0.6} \times 10^{23}\,\mathrm{cm^{-2}}$) are not large enough to efficiently absorb X-ray photons. Reconciling the predicted X-ray fluxes with the current observational upper limits requires $N_{\rm H}\sim 5\times10^{23}\,\mathrm{cm^{-2}}$, approximately $2.5$ times the upper bound of the simulated $1\sigma$ interval.

Neither of these mechanisms is currently included in our simulations. The \citetalias{Valentini:2021} subgrid model limits BH accretion to the Eddington rate and therefore does not explore the super-Eddington regime. It is worthwhile to notice that, on the one hand, only mild super-Eddington rates \citep[$1.4 < \lambda_{\rm EDD} < 4$;][]{Pacucci_2024} are required for having steep X-ray spectra; on the other hand, the \citetalias{Valentini:2021} simulations predict that, in the redshift range $6<z<7$, the SMBHs spend a substantial fraction of time accreting close to the Eddington limit (see their Fig. 10).  Incorporating physically motivated super-Eddington accretion models would therefore represent a natural extension of our framework. Although this may appear in tension with the sub-Eddington ratios inferred for many observed LRDs, this discrepancy may be alleviated by the reduced radiative efficiency expected for $\dot{M}/\dot{M}_{\rm Edd}>1$, where the emergent luminosity no longer scales linearly with the mass accretion rate because of photon trapping within the accretion flow. This would allow rapidly accreting black holes to appear only moderately super- or even sub-Eddington when their luminosities are interpreted assuming the standard thin-disk radiative efficiency. 

Likewise, while the sub-parsec gas structures responsible for Compton-thick obscuration cannot be resolved in cosmological simulations, their radiative effects could be incorporated in a physically motivated way by adopting intrinsic AGN SED templates already filtered through large gas column densities. Such an approach would allow us to quantify how much additional dust attenuation is required to simultaneously reproduce the observed X-ray weakness and the optical V-shaped continuum of LRDs, while remaining fully consistent with the large-scale gas and dust distribution predicted by the simulations. 

Accounting for both super-Eddington accretion and a distribution of gas column densities may help explain the observed diversity in X-ray properties, including rare X-ray detected LRDs \citep{kocevski_rise_2024} and candidate transitional sources \citep{hviding2026xraydotexoticdust}.

\subsection{Evolved stellar populations and extreme Balmer breaks} 
Our simulations successfully reproduce the Balmer break strengths of the majority of LRDs, yielding $F_{4000}/F_{3600}\sim1$--2 through stellar populations alone, without requiring post-starburst ages. However, they do not include highly evolved stellar populations ($t_\star \sim 0.3$--$1\,\mathrm{Gyr}$), and therefore are not designed to reproduce stronger stellar-driven Balmer breaks, which can reach $F_{4000}/F_{3600}\sim2$--2.5. Incorporating such evolved populations in future simulations will enable a more direct comparison with observations and test whether these systems can be reproduced while remaining consistent with the rest-frame UV colours required for LRD selection. 

Likewise, our radiative transfer calculations do not include continuum absorption and reprocessing by dense screens of excited $n=2$ neutral hydrogen, which have recently been proposed to explain the most extreme Balmer discontinuities observed in objects such as The Cliff and MOM-BH*1 \citep[e.g.,][]{inayoshi_extremely_2025, matthee2026engineflowslittlered}. Because these absorbing layers also reshape the rest-frame UV and optical continuum, they are expected to affect the location of galaxies within the photometric LRD selection space. Incorporating both highly evolved stellar populations and radiative transfer through dense neutral-hydrogen screens will therefore be an important step toward testing whether the full diversity of Balmer break strengths observed in the LRD population can be reproduced within a self-consistent framework.   

\subsection{The low metallicities inferred for LRDs} 
The mass-weighted gas metallicity within $0.1\,r_{\rm vir}$ of our host halos ($Z_{\rm gas}\sim0.5$--$1\,Z_\odot$) reflects the chemically enriched environments expected for massive $z\sim6$ quasar hosts in the \citetalias{Valentini:2021} simulations. This appears higher than the sub-solar gas metallicities ($Z_{\rm gas}\sim0.05$--$0.15\,Z_\odot$) inferred for many observed LRDs from nebular emission-line diagnostics \citep[e.g.,][]{greene_uncover_2024, Killi_2024, Tripodi_2025, Torralba_2026}. Furthermore, recent studies suggest that low gas metallicities may represent a defining property of the LRD population over a broad redshift range \citep{nikopoulos2026metallicitieslittlereddot}.

However, a direct comparison between the simulated and inferred metallicities is not yet possible. The metallicities quoted for our simulations are mass-weighted quantities, whereas the observational estimates are derived from emission-line diagnostics that depend not only on the gas metallicity but also on its density, temperature, ionisation state, therefore, on the incident radiation field. A meaningful comparison therefore requires self-consistent photoionisation calculations (e.g. using \textsc{Cloudy}) performed on the simulated gas distribution, combining the local physical conditions predicted by the simulations with the radiation field obtained from the radiative transfer calculations. Indeed, recent analyses \citep[e.g.][]{2026Helton} have shown that the metallicities inferred from standard strong-line diagnostics may differ significantly from those obtained through detailed photoionisation modelling of the same spectra, reflecting the strong dependence of emission-line ratios on the underlying physical conditions of the gas. This reinforces the need for dedicated \textsc{Cloudy} calculations on our simulated galaxies before drawing quantitative conclusions about possible discrepancies in metallicity. Such an analysis would allow synthetic emission-line spectra to be constructed and analysed using the same diagnostics adopted for observed LRDs, providing a much more robust comparison between simulations and observations.

\section{Summary and conclusions} \label{summary}
We have investigated whether the characteristic properties of Little Red Dots, namely rest-optical compactness and a spectral V-shape, can arise naturally within cosmological simulations of massive AGN-host galaxies. We performed 3D dust continuum radiative transfer calculations using \texttt{SKIRTv8} to post-process the cosmological zoom-in simulations of galaxy and supermassive black hole (SMBH) formation presented in \citetalias{Valentini:2021}. Specifically, we analyzed snapshots at $z=6-9$ from two simulation suites: \emph{AGNfid}, which includes SMBH formation, accretion, and feedback; \emph{SFonly}, which only includes the physics of star formation. We further explored the impact of varying the dust-to-metal ratio ($f_{\rm d}$), grain size distribution and composition, and the intrinsic AGN spectral energy distribution. We developed a mock observation pipeline to convert the radiative transfer outputs into synthetic \textit{JWST}/NIRCam photometric maps. Focusing on the brightest sources within the central $5\,\text{kpc}$ region of the main halos, we applied the photometric selection criteria of \citetalias{kocevski_rise_2024} alongside the F444W compactness criterion of \citealt{Labbe_2025} to identify LRD candidates. 

Our main findings can be summarised as follows:
\begin{itemize}
    \item[$\bullet$] 
    
    Both the \emph{AGNfid} and \emph{SFonly} populations can exhibit the typical "V-shape" of LRDs. However, only systems containing an accreting BH simultaneously satisfy the photometric and compactness criteria used to identify LRDs. Within our simulations, an AGN is therefore required to reproduce the combination of red optical colours and compact rest-frame optical morphology characteristic of the observed population.
    
    \item[$\bullet$] Mock NIRCam observations reveal extended rest-frame UV structures, including host galaxy emission, merging companions, and stellar clumps, surrounding a compact red core. This morphology arises because the host remains visible in the UV while the central AGN dominates at longer wavelengths. These features closely resemble those observed in a substantial fraction of LRDs, with extended rest-UV emission reported in $\sim$40\% of the sample \citep{Rinaldi_2025, Baggen_2026}.
    
    \item[$\bullet$] The simulated LRDs reproduce the UV-bright portion of the observed population ($-22.5 < M_{\rm UV} < -19.5$) and exhibit moderate V-band attenuation ($A_{\rm V} < 3.5$). Their stellar populations produce moderate Balmer breaks, $F_{4000}/F_{3600}\sim1$--$2$, consistent with approximately $40$--$50\%$ of observed LRDs. The strongest observed discontinuities require additional ingredients, such as more evolved stellar populations or absorption and reprocessing by dense excited neutral hydrogen.
    
    %\item[$\bullet$] The normalised SED shapes of \emph{AGNfid} sources closely match the observed average LRD spectrum from \citealt{Perez2024}, whereas the \emph{SFonly} models exhibit a rapid drop-off in the rest-frame near to mid-infrared range (Fig. \ref{fig_obs:3}). 
    
    %\item[$\bullet$] Enforcing FIR non-detection upper limits on compact \emph{AGNfid} SEDs (Fig. \ref{fig_obs:4}) reveals that only objects with relatively low total dust masses ($M_{\rm d} \sim 1.6 \times 10^6\,M_{\odot}$) comply with empirical submm constraints \citep{Casey_2025}.
    
    \item[$\bullet$] The simulated LRDs host SMBHs that lie systematically above the local $M_{\rm BH}-M_\star$ relations. In contrast, some compact, non-V-shaped \emph{AGNfid} objects lie closer to the local scaling relations, mirroring the spectroscopically inferred differences between LRD and Little Blue Dot (LBD) populations. The simulations also reproduce the continuous relation between $L_{5100}$ and $\beta_{\rm opt}$ inferred for LRDs and LBDs. Together, these results support an interpretation in which the two populations are not intrinsically distinct, but occupy different regions of a continuous distribution governed by the relative AGN contribution and line-of-sight attenuation.
    
    \item[$\bullet$] The simulated LRDs span $0.05 < \lambda_{\rm Edd} < 0.5$, broadly consistent with the majority of current observational estimates after accounting for their systematic uncertainties. Sources undergoing super-Eddington accretion remain outside the parameter space explored here because the adopted subgrid model caps accretion at the Eddington limit.
    
    \item[$\bullet$] Spectral and photometric decompositions show that the V-shaped continuum results from the different spatial distributions and attenuation experienced by the stellar and AGN components. The steep rest-optical slope ($\beta_{\rm opt}>0$) is produced by an intrinsically red, heavily dust-attenuated AGN, while the less obscured stellar component dominates the rest-UV emission ($\beta_{\rm UV}<-2.8$). Consistently, the increasing contribution of the unresolved AGN toward longer wavelengths naturally drives the high spatial compactness. This coupling between red rest-optical colours and compact morphology agrees with the observations of the RUBIES survey \citep{Hviding_2025}.
    %The V-shaped continuum results from the different spatial distributions and attenuation experienced by the stellar and AGN components. The relatively unobscured host galaxy dominates the blue rest-frame UV continuum, whereas the dust-attenuated AGN emerges redward of the Balmer break and dominates the rest-frame optical. Because this optical component is unresolved, the same mechanism naturally links red optical colours to compact morphology.
    
    %, providing a natural explanation for LRD continuum SEDs without invoking post-starburst stellar populations \citep[e.g.,][]{labbe_unambiguous_2024, Wang_2024, ma_uncover_2024} or dense screens of excited $n=2$ neutral hydrogen \citep[e.g.,][]{inayoshi_extremely_2025, matthee2026engineflowslittlered}.
\end{itemize}

Taken together, our results indicate that many of the defining properties of LRDs can arise naturally in massive high-redshift AGN-host galaxies. In this picture, LRDs do not constitute a fundamentally distinct class of objects: their characteristic colours and morphologies emerge from the wavelength-dependent interplay between extended stellar emission, an obscured central engine, and line-of-sight dust geometry.

The present framework does not yet reproduce the full diversity of the observed population. This provides a clear roadmap for future work, highlighting the need to incorporate additional physical ingredients. In particular, our model does not include radiation reprocessing by a parsec-scale dusty torus, super-Eddington accretion, or absorption by compact dust-free gas columns, and it lacks the evolved stellar populations or excited neutral-hydrogen screens that may produce the strongest Balmer breaks. Moreover, quantitative comparisons with metallicity estimates require self-consistent photoionisation modelling of the simulated gas. Incorporating these ingredients will test whether a unified AGN-host framework can simultaneously explain the X-ray, infrared, and optical diversity of LRDs and their possible evolution into, or connection with, the broader high-redshift AGN population.
\begin{acknowledgements} 
The authors thank Roberto Maiolino for valuable comments and insightful discussions on the results and limitations of this work. SC acknowledges support from PNRR funds. SG acknowledges support from the PRIN 2022 project (2022TKPB2P), titled: "BIG-z: Building the Giants: accretion, feedback and assembly in z>6 quasars". MV is supported by the Fondazione ICSC National Recovery and Resilience Plan (PNRR), Project ID CN-00000013 "Italian Research Center on High-Performance Computing, Big Data and Quantum Computing" funded by MUR - Next Generation EU. MV also acknowledges partial support from the INFN Indark Grant. The authors acknowledge the use of {\it ChatGPT} and {\it Google Gemini} to improve the readability of the text. The scientific content and conclusions remain entirely the authors' responsibility. This work was supported by the Open Access Publishing Fund of the Scuola Normale Superiore.
\end{acknowledgements}

\bibliographystyle{aa_url}
\bibliography{biblio}
%\bsp
\appendix
\section{NIRCam Maps} \label{app:allmaps}
In Fig.~\ref{fig:allmaps}, we present the mock \emph{JWST}/NIRCam image stamps across all six selection filters (\textsf{F115W}--\textsf{F444W}) for the remaining LRDs in our sample, complementing the representative cases shown in Sec.~\ref{res:morph}. Across these maps, the sources consistently maintain a compact morphology in the rest-frame optical filters (\textsf{F277W}--\textsf{F444W}). In contrast, the rest-frame UV filters (\textsf{F115W}--\textsf{F200W}) display varying degrees of the morphological structures (described in Sec.~\ref{res:morph}), which range from compact cores and stellar clumps to merging companions or diffuse host galaxy emission.
\begin{figure*}[h]
    \centering
    \includegraphics[width = 0.99\linewidth]{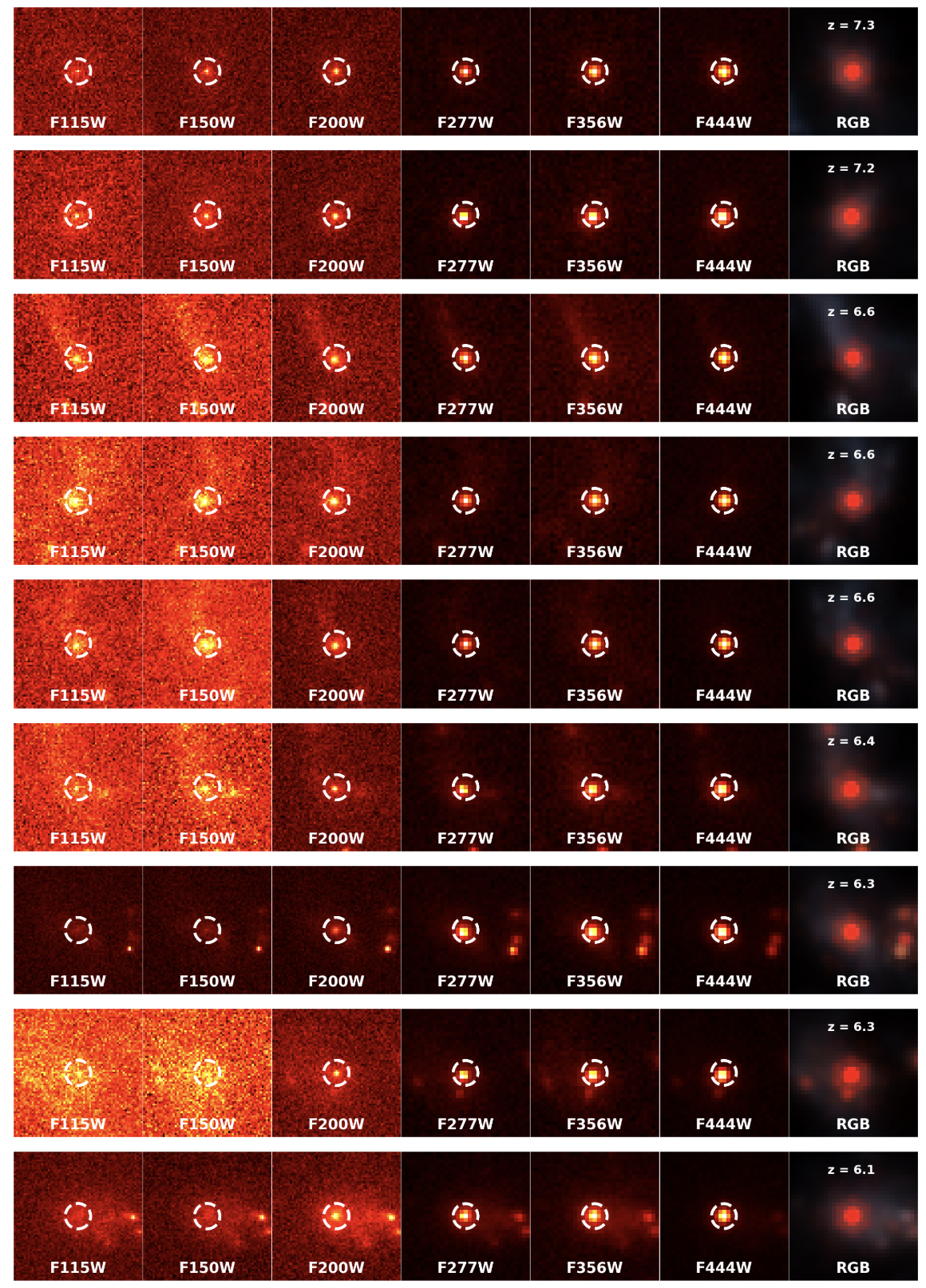}
\end{figure*}
\begin{figure*}[h]
    \centering
    \includegraphics[width = 0.99\linewidth]{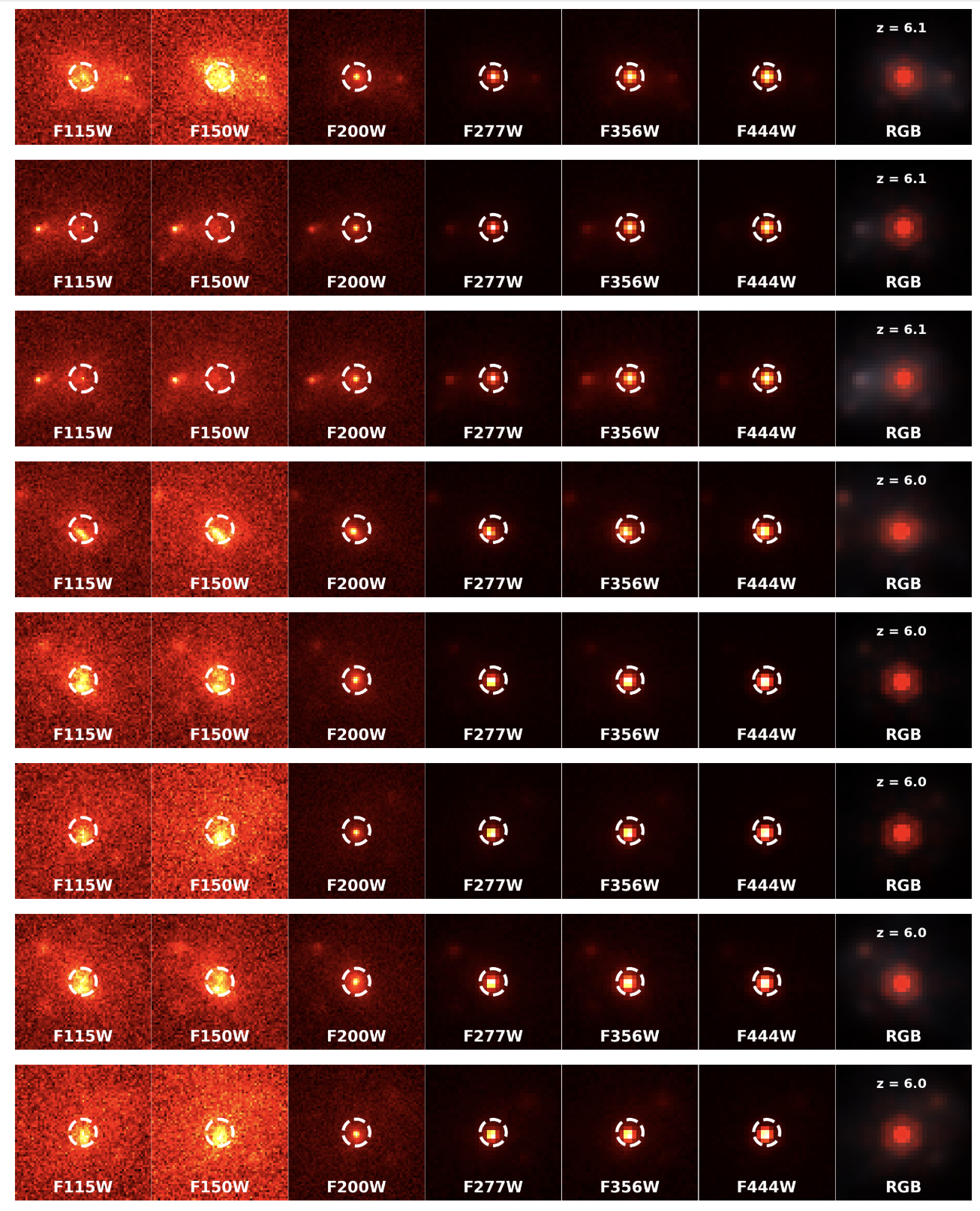}
    \caption{Mock \emph{JWST}/NIRCam image stamps and corresponding three-colour RGB composites generated using the post-processing procedure described in Sec.~\ref{mocks} for all simulated LRDs selected in Sec.~\ref{res:LRDphoto}, except for those already shown in Sec.~\ref{res:morph}. The redshift of each source is indicated in the RGB panel. Each  panel shows a $2'' \times 2''$ field of view, centered on the most luminous source within the central massive halo in the NIRCam F444W band. The dashed white circle overplotted on single-filter each panel represents the $D_{\rm ap} = 0.4''$ diameter photometric aperture used for flux extraction. The rightmost column shows RGB composites combining the F444W (R), F200W (G), and F115W (B) filters.}
    \label{fig:allmaps}
\end{figure*}
\vspace{-0.2cm}
\section{Mid-Infrared Continuum} \label{res:MIRI}
\begin{figure}[h]
    \centering
    \includegraphics[width = \linewidth]{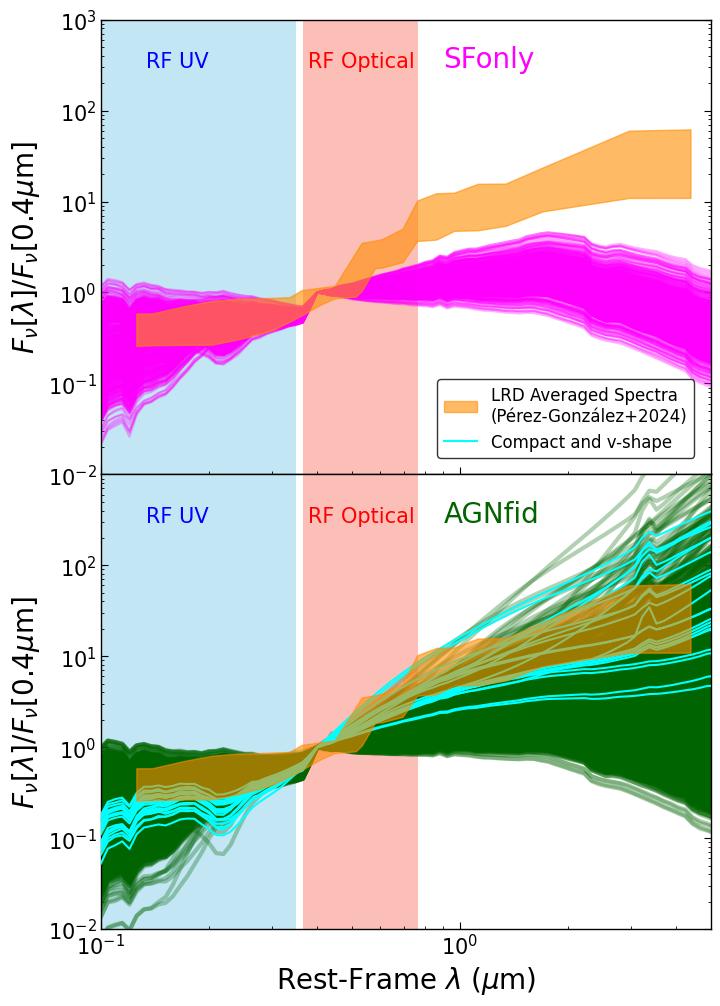}
    \caption{Synthetic rest-frame SEDs for the \emph{SFonly} (\textbf{top panel} in green) and \emph{AGNfid} (\textbf{bottom panel} in magenta) simulation suites, normalised at rest-frame $0.4\,\mu\text{m}$. The compact and photometrically selected LRD sources are highlighted in cyan. For empirical comparison, the orange shaded region represents the average observed LRD spectrum from \citealt{Perez2024}. Vertical shaded bands denote the rest-frame (RF) UV (light blue) and RF optical (light red) wavelength windows used to evaluate the spectral indices ($\beta_{\text{UV}}$ and $\beta_{\text{opt}}$) for the selection criteria.}
    \label{fig_obs:3}
\end{figure}

Fig.~\ref{fig_obs:3} shows the rest-frame SEDs of our simulated sources, normalised at $0.4~\mu\mathrm{m}$ to allow a direct comparison with the empirical average LRD spectrum from \citealt{Perez2024}. All of the \emph{SFonly} SEDs decline rapidly beyond the rest-frame optical, failing to reproduce the pronounced NIR/MIR excess observed in LRDs. Instead, many of the \emph{AGNfid} sources reproduce the observed average LRD spectrum remarkably well. 

This comparison should be interpreted with caution. For what concerns the \emph{SFonly} simulations, although our systems can reach star formation rates $\text{SFR} \sim 100~\rm M_{\odot} ~yr^{-1}$, and stellar masses
$M_\star\sim 10^{10}\,M_\odot$, they are distributed over a radius of $\sim 5~\rm kpc$. Their star-formation surface densities ($\Sigma_{\rm SFR}\sim1\,M_\odot\,{\rm yr}^{-1}\,{\rm kpc}^{-2}$) and stellar surface densities ($\Sigma_\star\sim 10^8\,M_\odot\,{\rm kpc}^{-2}$) are therefore far below those required by the compact dusty starburst scenarios to explain LRDs \citep[$\Sigma_{\rm SFR}\gtrsim 10^2-10^3\,M_\odot\,{\rm yr}^{-1}\,{\rm kpc}^{-2}$, and $\Sigma_\star\gtrsim 10^{11}\,M_\odot\,{\rm kpc}^{-2}$, respectively;][]{Barro_2023, baggen_small_2024, Perez2024}.
For what concerns the \emph{AGNfid} simulations, the NIR/MIR excess in observed sightlines simply reflects the attenuated AGN continuum. In our radiative-transfer calculations, the AGN SED does not include emission from a conventional hot dusty torus (see Sec.~\ref{sec:SEDs}). Therefore, while our results support the observational inference that LRDs lack the strong hot-dust emission predicted by standard AGN templates \citep{Williams2023, Akins2024,  Wang_2025, Li_2025}, they cannot be used to assess whether such a torus is physically absent. Addressing this question requires dedicated torus modelling and cosmological hydrodynamical simulations with a much higher resolution, which we defer to future work. %\SG{Saksham, could you please overplot the intrinsic, normalised continuum as well in Fig. 7? I think that at least we can say that the shape, after dust reddening, is fine.}

\section{Far-Infrared Upper Limits} \label{res:fir_limits}
\begin{figure}[h]
    \centering
    \includegraphics[width = \linewidth]{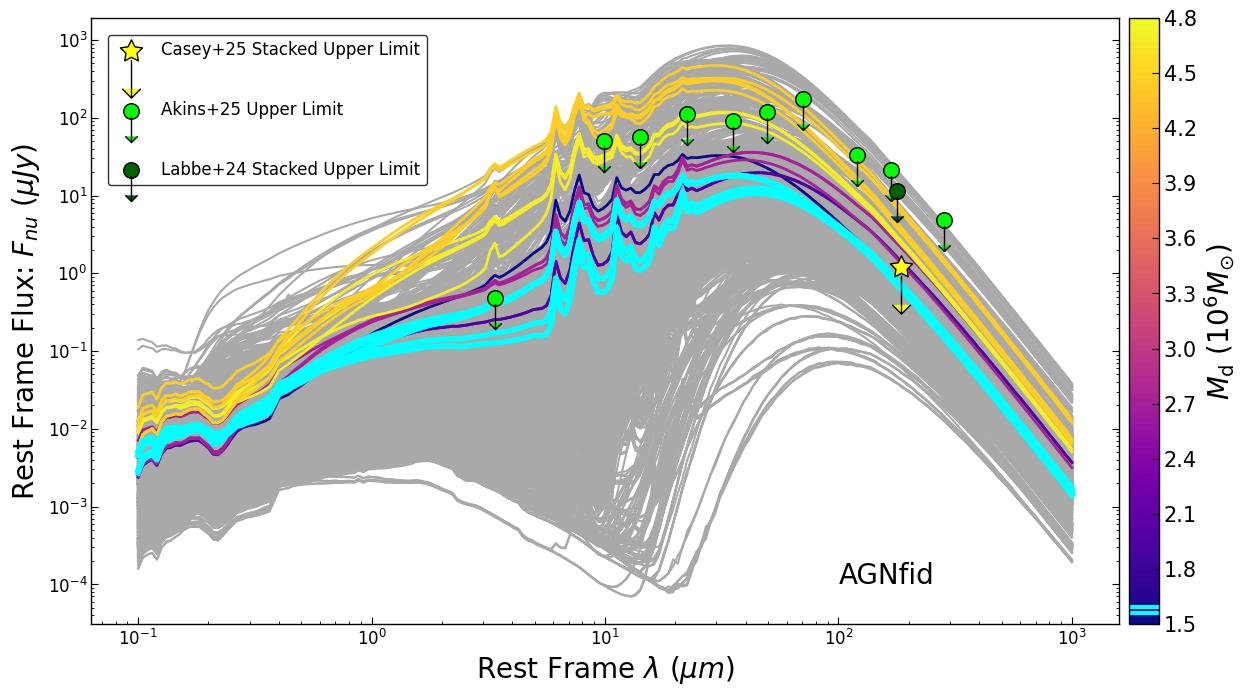}
    \caption{Synthetic rest-frame \emph{AGNfid} SEDs evaluated against empirical FIR and submillimeter upper limits. All \emph{AGNfid} sources are plotted in grey, while the morphologically compact and V-shape selected LRD candidates are colour-coded by their dust mass ($M_{\text{d}}$ evaluated within $0.1\,r_{\text{vir}}$, where $r_{\text{vir}}$ is the virial radius of the host halo). Cyan curves highlight the surviving sightlines that remain consistent with the $5\sigma$ empirical constraints: stacked non-detections from \citealt{Akins2024} (ALMA, SCUBA-2, Spitzer/MIPS, and Herschel/PACS/SPIRE; green circles), ALMA Band 6 constraints from \citealt{Labbe_2025} (dark green circle), and constraints from \citealt{Casey_2025} (yellow star).}
    \label{fig_obs:4}
\end{figure}

In Fig. \ref{fig_obs:4}, we compare the rest-frame SEDs derived from our RT calculations with available rest-frame FIR and submillimeter upper limits from the literature \citep{Labbe_2025, Akins2024, Casey_2025}.
%Evaluating these distributions in the rest-frame eliminates the dependence of redshift, allowing for a direct assessment of how submillimeter non-detections constrain the underlying SED templates. 
%\emph{AGNfid} sources that fail the structural compactness are not taken into consideration, because our synthetic photometry is extracted within a fixed $D_{\rm ap} = 0.4''$ aperture, making it methodologically self-consistent to isolate the compact subpopulation where the vast majority of the total integrated flux originates from the central core.
A significant fraction of the simulated SEDs, particularly those with dust reservoirs $M_{\rm d} \gtrsim 2\times 10^6~\mathrm{M}_\odot$, exceed the available FIR and submillimeter upper limits, consistent with the empirical constraints of $M_{\rm d} < 10^6~\mathrm{M}_\odot$ reported by \citealt{Casey_2025}. The corresponding thermal dust emission peaks between rest-frame $30~\mu\mathrm{m}$ and $100~\mu\mathrm{m}$, coinciding with the wavelength range where the current FIR and submillimeter observations provide the strongest constraints. Consequently, only systems with relatively low dust masses ($M_{\rm d} \sim 1.6 \times 10^6~\mathrm{M}_\odot$, shown in cyan) remain consistent with all the available FIR and submillimeter upper limits. Incorporating an unresolved nuclear dusty torus would further refine these constraints, as central dust absorption converts a portion of the AGN emission into warmer MIR radiation, thereby reducing the thermal heating of the large-scale dust reservoir.
%A significant fraction of these simulated SEDs—specifically those with high dust reservoirs ($M_{\rm d} \gtrsim 0.9 \times 10^6\,\text{M}_\odot$, clearly violate the stacked submillimeter upper limits, a result that is highly consistent with the empirical exclusions outlined by \citealt{Casey_2025}. The simulated dust peak consistently materializes between rest-frame $30\,\mu\text{m}$ and $100\,\mu\text{m}$, directly hitting the maximum sensitivity regimes of these submillimeter frameworks. A substantial subset of our colour-selected LRDs overlap with or exceed the FIR constraints, and only systems possessing low dust masses ($M_{\rm d} \lesssim 0.4 \times 10^6\,\text{M}_\odot$, shown in purple) survive these joint submillimeter boundaries.

The strength and wavelength of the FIR peak are determined not only by the dust masses but also by the amount of absorbed UV radiation. Since dust re-emission is constrained by energy conservation, intrinsically UV-luminous sources produce brighter FIR emission and hotter dust temperatures, shifting the thermal peak toward shorter wavelengths even for relatively modest dust masses. Because our sample focuses on the central, most massive galaxies hosting rapidly accreting black holes, the selected sources are representative of the brightest LRDs observed in the UV (see Sec.~\ref{sec:uv}). Consequently, even systems with relatively small dust reservoirs can produce FIR emission exceeding the current observational limits. 
Conversely, the observed LRDs may remain undetected in the FIR despite having comparable dust masses simply because they are intrinsically less UV luminous. %In Appendix \ref{app:intrinsic}, we explore how reducing the intrinsic luminosity while preserving the observed photometric properties affects these conclusions.

Among the available observational constraints, the stacked ALMA Band 6 ($1.2\,\mathrm{mm}$) upper limit of \citealt{Casey_2025}, and the \emph{Spitzer}/MIPS $24~\mu\mathrm{m}$ limit compiled by \citealt{Akins2024} provide the strongest constraints on our simulated SEDs. The ALMA observations probe the Rayleigh-Jeans tail of the cold dust emission, while the MIPS $24\,\mu\mathrm{m}$ band samples the transition between the near- and far-infrared ($\sim 3~\mu\mathrm{m}$ rest-frame at $z\sim 6$), where the AGN and dust contributions become comparable.

\section{X-ray absorption}\label{compton_thick} 
We attenuate the intrinsic power-law spectrum, by a gas column density $N_{\rm H}$, such that:
\begin{equation}
\frac{dN}{dE} = A E^{-\Gamma}
\exp\left[-N_{\rm H}\sigma_{\rm eff}(E)\right],
\end{equation}
where $\Gamma=1.7$ and $\sigma_{\rm eff}$ is the effective cross-section given by \citep{Clark_1998}:
\begin{equation}
\sigma_{\rm eff}(E)=\sigma_{\rm ph}(E)+\sigma_{\rm T}{\mu_e}.
\end{equation}
Here, $\sigma_{\rm ph}$ is the photoelectric absorption \citep{Maloney_1996}\footnote{This power-law approximation provides a suitable order-of-magnitude estimate, though it neglects metal abundance variations, the Fe K-edge feature at $\sim 7.1\,\mathrm{keV}$, and Compton scattering or reflection.}:
\begin{equation}
\sigma_{\rm ph}(E)=3\times10^{-22}
\left(
\frac{E}{1,\mathrm{keV}}
\right)^{-8/3}\mathrm{cm^2},
\end{equation}
where $\sigma_{\rm T}=6.65\times10^{-25}\,\mathrm{cm^2}$ is the Thomson cross-section, and $\mu_e\simeq1.2$ accounts for the number of free electrons per unit gas column for standard cosmic abundances. 

The attenuated hard X-ray luminosity in the energy interval $[E_1,E_2]=[2~\rm keV,10~\rm keV]$ is then
\begin{equation}
L_{\rm X,abs}(N_{\rm H})=C
\int_{E_1}^{E_2}
E^{1-\Gamma}
\exp\left[-N_{\rm H}\sigma_{\rm eff}(E)\right],dE .
\end{equation}

Rather than explicitly determining the normalisation $C$, we compute the attenuation relative to the intrinsic luminosity. The transmitted fraction in the $[E_1,E_2]$ band is therefore
\begin{equation}
f_{\rm transm}=\frac{
\displaystyle
\int_{E_1}^{E_2}
E^{1-\Gamma}
\exp\left[-N_{\rm H}\sigma_{\rm eff}(E)\right],dE
}{
\displaystyle
\int_{E_1}^{E_2}
E^{1-\Gamma},dE
},
\end{equation}
such that
\begin{equation}
L_{X,\rm obs}=f_{\rm transm}~L_{X,\rm intr}.
\end{equation}
where, $L_{X,\rm intr}$ is derived from $L_{\rm bol}$ using the bolometric corrections of \citealt{Duras2020}.

%Additionally, our modelling demonstrates that the ALMA Band 6 $ 1.2\,\text{mm}$) stacked upper limit from \citealt{Casey_2025} and the \emph{Spitzer}/MIPS $24\,\mu\text{m}$ constraint compiled by \citealt{Akins2024} provide the most stringent bottlenecks for the LRD SED shape. While ALMA limits the cold dust mass component on the Rayleigh-Jeans tail, the MIPS $24\,\mu\text{m}$ filter directly probes the rest-frame mid-to-far-IR transition ($\sim 3\,\mu\text{m}$ at $z\sim6$), making it uniquely sensitive to the warm dust component fueled by active accretion. We note, however, a critical observational caveat: several of the simulated compact objects that formally violate these limits might still evade detection in real shallow surveys. Due to the large beam sizes of these instruments—characterised by a Full Width at Half Maximum (FWHM) of $\sim 0.8''$ for ALMA Band 6 and PSF of $\sim 6''$ for MIPS—the total flux can be diluted by spatial beam smoothing, or may require deeper integrations to be robustly disentangled from nearby line-of-sight structures.

% Shhould we make this section more like that these objects would not be detected, or about what the already set upper lomits tell us about the shape of the SEDs. If the latter, is the plot shown here even relevant?
%\section{Reducing Intrinsic Emission} \label{app:intrinsic}
\label{lastpage}
\end{document}